\documentclass[11pt]{article}
\usepackage{authblk}
\usepackage[a4paper]{geometry}
\usepackage{graphicx}
\usepackage[textwidth=8em,textsize=small]{todonotes}
\usepackage{amsmath}
\usepackage{hyperref}
\usepackage{textcomp}
\usepackage{epstopdf}
\usepackage{setspace}
\usepackage{multirow}
\usepackage{tabularx}
\usepackage{longtable} 
\usepackage{pdflscape}
\usepackage{bm} 

\makeatletter 
\renewcommand\@biblabel[1]{#1.} 
\makeatother

\title{The critical role of  the interaction potential and simulation
       protocol for the structural and mechanical properties of sodosilicate glasses
       }
\date{}
\author{Zhen Zhang$^{(1)}$}
\author{Simona Ispas$^{(1)}$\thanks{Corresponding author: \texttt{simona.ispas@umontpellier.fr}}}
\author{Walter Kob$^{(1),(2)}$\thanks{Corresponding author: \texttt{walter.kob@umontpellier.fr}}}
\affil{$^{(1)}$ Laboratoire Charles Coulomb,
University of Montpellier, CNRS, \\
F-34095 Montpellier, France\\
$^{(2)}$ Institut Universitaire de France
}

\begin{document}

\maketitle

\begin{abstract}  
We compare the ability of various interaction potentials to predict
the structural and mechanical properties of silica and sodium silicate
glasses. While most structural quantities show a relatively mild
dependence on the potential used, the mechanical properties such as the
failure stress and strain as well as the elastic moduli depend very
strongly on the potential, once finite size effects have been taken
into account. We find that to avoid such finite size effects, samples of
at least 75,000 atoms are needed. Finally we probe how the simulation
ensemble influences the fracture properties of the glasses and conclude
that fracture simulations should be carried out in the constant pressure
ensemble.  \\

\end{abstract}

\section{Introduction}

Computer simulations have by now been established to be a highly valuable
tool to get insight into the properties of complex systems such as
liquids and glasses~\cite{kob_computer_1999,massobrio2015molecular, kob_first-principles_2016}. The most important ingredient
for such simulations is the interaction potential between the particles
since its accuracy is crucial for obtaining a reliable description of the
properties of the material on the micro-and macroscopic scale. There are
two possibilities to describe these interactions: {\it ab initio} or using
effective potentials (also called ``classical simulations''). 
In the first approach the interactions
are calculated directly from the electronic structure of the system, a
procedure that is very accurate (although not exact) but computationally
very expensive~\cite{marx_ab_2009}. As a consequence only relatively small systems
can be simulated, i.e., typically less than $10^3$ particles, over a short
time (less than 10~ns). In contrast to this, effective potentials allow
to access significantly larger systems and longer times: $10^7$ atoms and
tens of micro-seconds. This advantage comes at the cost of a deteriorated
accuracy of the potentials and hence the predicted material properties
are less reliable. Despite this drawback many numerical investigations
are done with effective potentials since for many studies one needs
to have systems that are relatively large, e.g.~to probe mechanical
properties, and cooling rates that are not too high. As a consequence
significant efforts have been taken to obtain effective potentials that
are reliable and for systems like silica one can find in the literature
dozens of potentials, see Ref.~\cite{sundararaman_new_2018} for an exhaustive list.
(In recent
times force fields are sometimes obtained from approaches called ``machine
learning''~\cite{behler_perspective_2016,liu_machine_2019}. Since at
present it is not clear to what extent these approaches give transferable
potentials we will here not discuss them further.)

It is obviously most important to know which ones of these potentials are
reliable and which ones are not. Therefore one can find several studies
in the literature in which the performance of various potentials are
compared~\cite{hemmatti_ir_1997,hemmatti_comparison_2000,soules_silica_2011,cowen_force_2015,yuan_local_2001,bauchy_structural_2014}.
Roughly speaking one finds that the structural properties, such as the
static structure factor, are relatively independent of the potential
considered, a result that is not surprising since the parameters of many
potentials have been optimized to reproduce the glass structure determined
in experiments. Dynamical quantities, such as the diffusion constant,
show a much stronger dependence on the potential used~\cite{hemmatti_comparison_2000}.
This result is related to the fact that dynamics depends not only on the
interaction energy close to the local minima of the potential energy landscape, which
governs the structural properties of the glass, but also on the local
barriers, i.e., quantities that are usually not taken into account when
the parameters of a potential are optimized. As an example we mention here
the work by Hemmatti and Angell who probed how the diffusion constant
of Si in SiO$_2$ depends on the potential and found that the predicted
values varied by orders of magnitude~\cite{hemmatti_comparison_2000}.

As mentioned above, the investigation of mechanical properties of glasses
are one of the important applications for simulations with effective
potentials.  Whether the potentials that are currently used to simulate
such systems are indeed reliable to probe these properties has, however,
not been tested so far.  The goal of the present work is therefore to
fill this gap by investigating the fracture behavior of glasses. In
parallel we will also study how the ensemble used for the simulation,
constant pressure or constant volume, influences the fracture.

For the present study we focus on the case of pure silica and the
sodo-silicate system Na$_2$O-3SiO$_2$, in the following denoted as NS3, since
these are two glass-forming systems that are representatives for many
oxide-glasses.  The paper is organized as follows: In Sec.~\ref{sec:sim}
we give the details of the investigated potentials and the simulation
procedure. In Sec.~\ref{sec:res} we present the results on the static,
dynamical and mechanical properties of the various glass samples. The
paper is concluded with a summary and a discussion in the last section.

\section{Simulation details}
\label{sec:sim}

\subsection{Interatomic potentials}
For the simulation of SiO$_2$ and Na$_2$O-3SiO$_2$ (NS3) we considered
four pair potentials, all of which have the same functional form given by

\begin{equation} 
\label{eq:potential}
V(r_{ij}) =  \frac{q_iq_je^2}{4\pi \epsilon_0 r_{ij}} + 
A_{ij}e^{-r_{ij}/B_{ij}} - \frac{C_{ij}}{r_{ij}^6} \quad ,
\end{equation}

\noindent
i.e., they are given by the sum of a Coulomb and Buckingham
term. (Here $ r_{ij} $ is the distance between atoms $i$ and $j$.)
Thus the difference between the potentials are just the values for
the various parameters of the potential, i.e., $q_i$, $A_{ij}$
etc. In particular we will consider the potentials proposed by
Habasaki and Okada (HO)~\cite{habasaki_molecular_1992}, by Teter
{\it et al.}~\cite{cormack_alkali_2002}, by Guillot and Sator
(GS)~\cite{guillot_computer_2007}, and by Sundararaman {\it et al.}
(SHIK)~\cite{sundararaman_new_2018,sundararaman_new_2019}.  For the case
of silica we also considered the potential by van Beest {\it et al.}
(BKS)~\cite{bks-prl1990} since it has been used in many previous studies
and found to be able to reproduce quite well many properties of real
silica~\cite{vollmayr_cooling-rate_1996,horbach_static_1999,luo2018molecular}.
The parameters of the various potentials used in the present work are
given in Table~\ref{tab:potentials}.

The presence of the Coulomb term in Eq.~(\ref{eq:potential})
makes the use of such potentials computationally expensive since
they have to be evaluated by means of approaches like the Ewald
summation~\cite{allen_computer_2017}. One possibility to avoid this
problem is to use the method proposed by Wolf {\it et al} in which the
Coulomb term is replaced by

\begin{equation}
\label{eq:wolf}
\frac{q_iq_je^2}{4\pi \epsilon_0 r_{ij}} \rightarrow \left\{\begin{array}{cc}
 \displaystyle \frac{q_iq_je^2}{4\pi\epsilon_0}
\left[\left(\frac{1}{r_{ij}}-\frac{1}{r_c}\right)+\frac{r_{ij}-r_c}{r_c^2}\right] & r < r_c\\
 0 & r \ge r_c \end{array} \right.
\end{equation}

\noindent
where $r_c$ is a cutoff
distance~\cite{wolf_exact_1999,carre2007amorphous}.
 In this form the potential becomes thus short ranged and hence computationally
much more efficient. For the case of the SHIK potential we have
compared the structural, dynamical, and mechanical properties of
the glass if the Coulomb term is replaced by the expression given by
Eq.~(\ref{eq:wolf}) and in the following we will show that there is no
significant difference. Hereafter we will denote the potential
with Eq.~(\ref{eq:wolf}) as SHIK and the one with the full Coulomb
interaction as SHIKc.

\begin{table}[htbp]
\center
 \begin{tabular}{llllll} 
  		\hline
Parameters	& BKS & GS     &Teter       &HO     &	SHIK  \\ \hline
  		$A_{\mathrm{SiO}}$   [eV]                               & 18003.7572 & 50306.4259  & 13702.9050 & 10631.4994      & 23107.8476   \\ 
  			$A_{\mathrm{OO}}$                             & 1388.7730  & 9022.8533   & 1844.7458  & 1742.1231       & 1120.5290    \\ 	
  		
  		$A_{\mathrm{SiSi}}$                               & 0.0000     & 0.0000      & 0.0000     & 865032008.0130    & 2797.9792    \\ 
  		
  		$A_{\mathrm{NaO}}$                               &            & 120304.5810 & 4383.7555  & 1854.3947     & 1127566.0 \\ 
  		
  		$A_{\mathrm{SiNa}}$                            &            & 0.0000      & 0.0000     & 81407.6194     & 495653.0  \\ 
  		
  		$A_{\mathrm{NaNa}}$                           &            & 0.0000      & 0.0000     & 2558.6814        & 1476.9000    \\ \hline

  		$B_{\mathrm{SiO}}$   [\AA$ $]                                & 0.2052     & 0.1610      & 0.1938     & 0.2085           & 0.1962       \\ 
  		$B_{\mathrm{OO}}$                             & 0.3623     & 0.2650      & 0.3436     & 0.3513       & 0.3457       \\ 
  		$B_{\mathrm{SiSi}}$                               & 1.0000     & 1.0000      & 1.0000     & 0.0657              & 0.2269       \\ 
  		
  		$B_{\mathrm{NaO}}$                                &            & 0.1700      & 0.2438     & 0.2603            & 0.1450       \\ 
  		
  		$B_{\mathrm{SiNa}}$                              &            & 1.0000      & 1.0000     & 0.1175           & 0.1847       \\ 
  		
  		$B_{\mathrm{NaNa}}$                              &            & 1.0000      & 1.0000     & 0.1692        & 0.2935       \\ \hline	
  		$C_{\mathrm{SiO}}$  [eV$\cdot$\AA$ ^{6} $]                                & 133.5381   & 46.2981     & 54.6810    & 69.9590         & 139.6948     \\ 
  			$C_{\mathrm{OO}}$                            & 175.0000   & 85.0927     & 192.5800   & 212.9333     & 26.1321      \\ 
  		
  		$C_{\mathrm{SiSi}}$                               & 0.0000     & 0.0000      & 0.0000     & 23.1044         & 0.0000       \\ 
  		
  		$C_{\mathrm{NaO}}$                                &            & 0.0000      & 30.7000    & 0.0000       & 40.5620      \\ 
  		
  		$C_{\mathrm{SiNa}}$                             &            & 0.0000      & 0.0000     & 0.0000       & 0.0000       \\ 
  		
  		$C_{\mathrm{NaNa}}$                              &            & 0.0000      & 0.0000     & 0.0000        & 0.0000       \\ \hline
  		
  		$q_{\mathrm{Si}}$  [e]                              & 2.4        & 1.89        & 2.4        & 2.4          & 1.7755       \\ 
  		
  		$q_{\mathrm{Na}}$ [e]                                &            & 0.4725      & 0.6        & 0.88       & 0.6018       \\
  		
  		$q_{\mathrm{O}}$ [e]                             & -1.2       & -0.945      & -1.2       &  -1.28$^\star$       &  -0.9328$^\star$  \\ \hline
  		$R_{cut}$ [\AA]      & 5.5/12.0       & 11.0/12.0      & 8.0/12.0       &  8.0/12.0       &  8.0/10.0  \\ \hline
\end{tabular} 	
\caption{
\label{tab:potentials} 
Parameters for the various potentials. The oxygen charges for the HO
and SHIK potentials, marked by a star, are given for the NS3 composition  (see text for
details). The oxygen charge used for silica when using the SHIK potential
is  $-q_{\mathrm{Si}}/2$. Also included are the cutoff distances $R_{cut}$
for the short range/long range parts of the potentials.
}
\end{table} 
  
Due to the van der Waals term in Eq.~(\ref{eq:potential}), the potentials
have a singularity at short distances. To prevent that particles fuse
together in a unphysical manner, we added a short range
repulsive term~\cite{sundararaman_new_2019}. This modification
does not affect at all the properties of the system at intermediate and
low temperatures and hence can be considered to be just a computational
trick to avoid this singularity.

Finally we mention that the charge of the oxygen atoms for the
HO and SHIK potentials depends on composition in order
to maintain charge neutrality of the system when the sodium
concentration is varied $q_{\mathrm
O}\displaystyle =\frac{(1-y)\, q_{{\mathrm {Si}}}+2y \, q_{\mathrm{
Na}}}{2-y}$, where $y$ is the Na$_2$O mole concentration~\cite{habasaki_molecular_1992,sundararaman_new_2019}. Thus the
oxygen charges reported in Table~\ref{tab:potentials} have been
calculated for $y=0.25$ which corresponds to the sodosilicate
composition studied in the present work.

\begin{figure}[htbp]
\centering
\includegraphics[width=16cm]{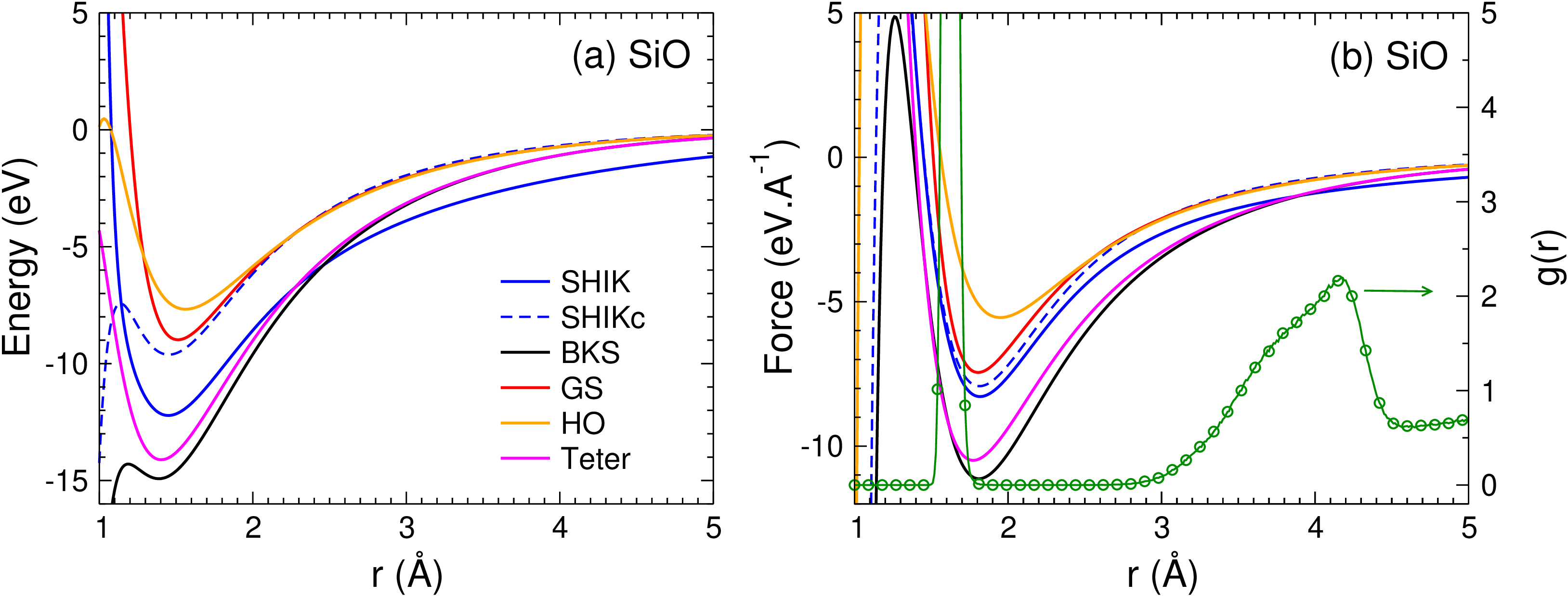}
\includegraphics[width=16cm]{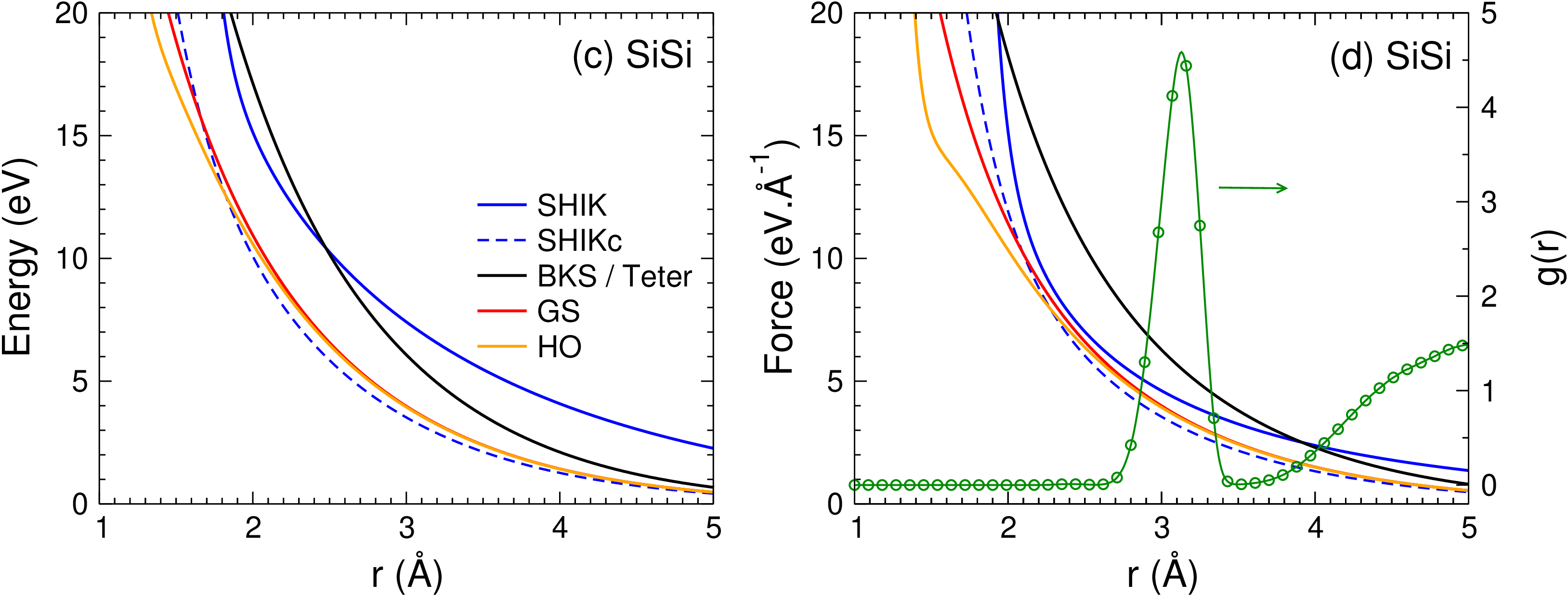}
\includegraphics[width=16cm]{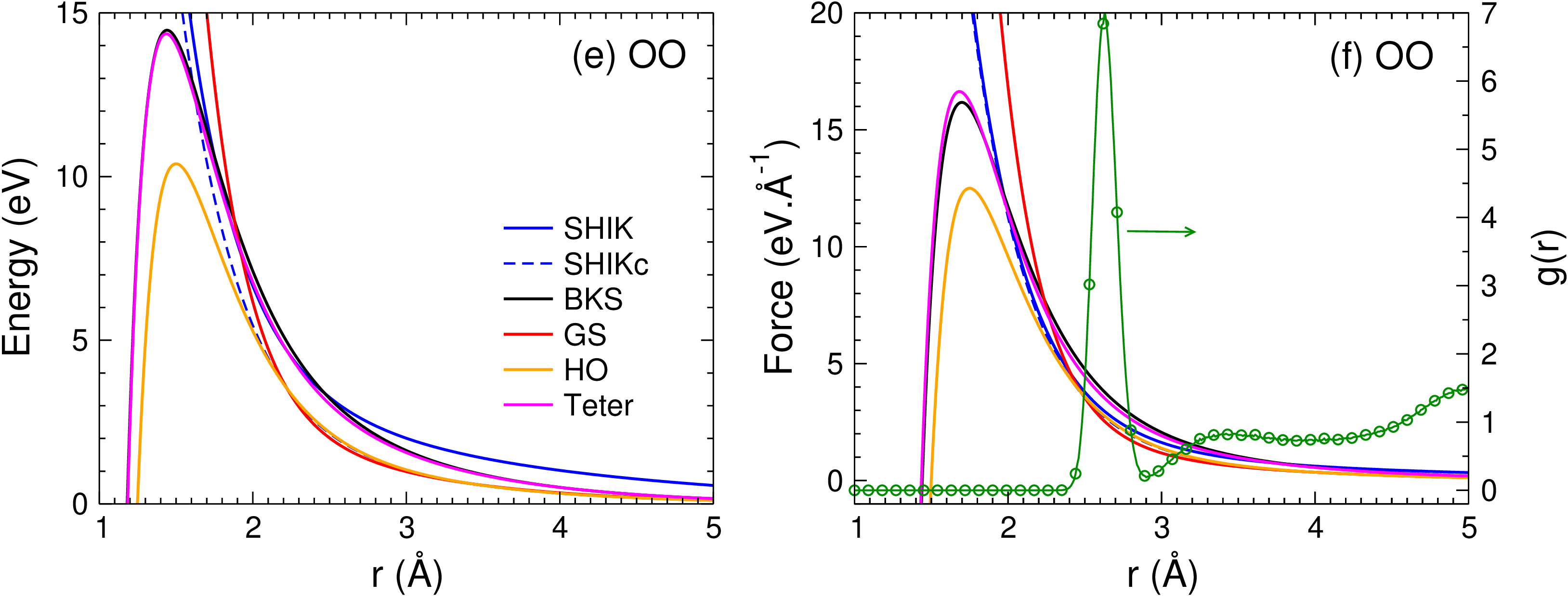}
\caption{
\label{fig:ns0-pot-force}  
Distance dependence of the various interaction potentials [panels
a), c), and e)] and their corresponding forces  [panels b), d), and
f)] for the SiO, SiSi and OO pairs. Also included in the latter panels 
are the corresponding pair correlation function as predicted by the SHIK potential at 300~K (symbols).
}
\end{figure}

In Fig.~\ref{fig:ns0-pot-force} we plot the potential energy and the
forces for the potentials considered. These graphs demonstrate that
the different potentials and forces depend strongly on the chosen set
of parameters which gives thus a first indication that the predicted
glass properties will depend on the potential used. Note that, in order
to show that we are indeed plotting the relevant range in distance,
we also include the radial distribution function $g_{\alpha\beta}(r)$,
with $\alpha,\beta \in \{{\rm Si, O}\}$ (green lines with symbols)  in
the graphs for the force. (These $g_{\alpha\beta}(r)$ are for the SHIK
potential, at 300~K, but the other potentials predict radial distribution
functions that are very similar.)

\subsection{Simulation procedure}

The systems we consider are pure silica and a sodium silicate with
composition Na$_2$O-3SiO$_2$  corresponding to a Na$_2$O molar
concentration of 25~\%. The glass samples were produced by using
the conventional  melt-quench method. We used cubic boxes (periodic
boundary conditions) that contained between 5000 and 600,000 particles,
which corresponds to sizes between 4 and 20~nm at room temperature.
For the long-range Coulomb interaction, the Wolf truncation method (see
Eq.~(\ref{eq:wolf})) was employed when using the SHIK potential
while for the other potentials this interaction was evaluated with the
particle-particle particle-mesh (PPPM) solver algorithm with an accuracy
of $5\times 10^{-5}$. As mentioned above, we also carried out simulations
for silica using the SHIK parameters for the short-range part of the
potential, and Coulomb interactions evaluated using PPPM algorithm,
and the corresponding data will be labeled SHIKc.

The samples were first equilibrated at a high temperature in the
canonical ensemble ($NVT$) using a fixed volume that corresponds
to the experimental value of the density of the glass at room
temperature~\cite{bansal_handbook_1986}.  These $NVT$ runs were done
at 3600~K for silica and  3000~K for NS3, both for about 300~ps, a time
that is sufficiently long to equilibrate the samples completely. These
liquids were subsequently equilibrated in the $NPT$ ensemble (constant
number of atoms, pressure, and temperature) at the same temperatures
and at zero pressure. The lengths of these $NPT$ runs depended on the
potential considered and were sufficiently long to equilibrate the
samples, i.e. the mean squared displacement of the particles
was more than 100~\AA$^2$ (see below).  For the GS potential, the $NPT$
equilibration of the NS3 liquid was done at 2100~K since for higher
temperatures the samples became unstable because they were above the boiling
point of the GS potential.

All simulations were carried out using the LAMMPS
software~\cite{plimpton_fast_1995} with a time step
of 1.6~fs and using a Nos\'e-Hoover thermostat and
barostat~\cite{nose_unified_1984,hoover_canonical_1985,hoover_constant-pressure_1986}.
After equilibration, the liquid samples were quenched to 300~K in the
$NPT$ ensemble at zero pressure. The cooling rate was 0.25~K/ps, i.e.~a
value that is relatively small for MD simulations of sodium silicate
glasses with a comparable system size. Previous simulation studies
have shown that this cooling rate is small enough so that the properties
of the system do not depend on the cooling rate in a significant
manner~\cite{vollmayr_cooling-rate_1996,lane_cooling_2015,li_cooling_2017}.
The glass samples at 300~K were then annealed in the $NPT$ ensemble for
160~ps before we started the mechanical tests.
 	Since room temperature is well below the glass transition temperature, there is no need to consider a longer thermalization run as the atomic configurations are practically frozen, and neither the density nor the structural properties change noticeably.

	The results presented in the following sections have been obtained by using for each potential only one sample per composition. For the structural and dynamical properties discussed in Sec. 3.1,  the samples contained 36,480 and 38,400~atoms for SiO$_2$ and NS3, respectively,  corresponding to box sizes around 8~nm. These system sizes are sufficiently large to make sample-to-sample fluctuations small.

	For a given composition and a given potential, the mechanical properties presented in  Subsec.~\ref{sec:tensile} are the average obtained from one sample put under uniaxial tensile strain in the three independent directions, i.e.~$x$, $y$, and $z$ directions, and the error bars  correspond to standard deviation.
	To calculate the stress-strain curve we increased the dimension of the box in one direction linearly in time, $L(t)$. The strain is then given by 
	\begin{equation} 
	\epsilon(t)=  \dot{\epsilon} t
	\end{equation}
	where $ \dot{\epsilon} $  is the strain rate.
	The stress tensor is obtained from the usual expression~\cite{thompson2009general}
	\begin{equation}
 {\bm \sigma}	 = \frac{1}{V} \sum_{i=1}^{N}\left[ m_i \mathbf v_i \otimes \mathbf v_i +\mathbf r_i \otimes \mathbf f_i\right]
	\end{equation}
	where $V$ and $N$ are the volume and the total number of atoms of the simulation box,  respectively,  while $m_i$  is the mass of atom $i$, and  $\mathbf v_i$, $\mathbf r_i$ and $\mathbf f_i$ are the velocity, position and force vector of atom $i$, respectively.
\\

\section{Results}
\label{sec:res}

\subsection{Structural and dynamical properties in the liquid and glassy states}
\label{sec:struct_dyn}

The results in the present subsection are all for a cubic system with
the size of the edge given by $L=8$~nm which corresponds to $N=36480$
and $N=38400$ atoms for SiO$_2$ and NS3, respectively. This size is
sufficiently large to avoid that dynamic and static quantities are
affected by finite size effects.

To investigate the dependence of the dynamics on the potential we have
calculated the mean squared displacement (MSD) of a tagged particle:

\begin{equation}
r_\alpha^2 (t) = \frac{1}{N_\alpha} \sum_{j=1}^{N_\alpha} 
\langle |{\bf r}_j(t)- {\bf r}_j(0)|^2 \rangle \quad ,
\label{eq:msd}
\end{equation}

\noindent
where $\alpha \in \{ {\rm Si, O, Na}\}$ and $N_\alpha$ is the number of
particles of species $\alpha$.

\begin{figure}[htbp]
\centering
\includegraphics[width=7.5cm]{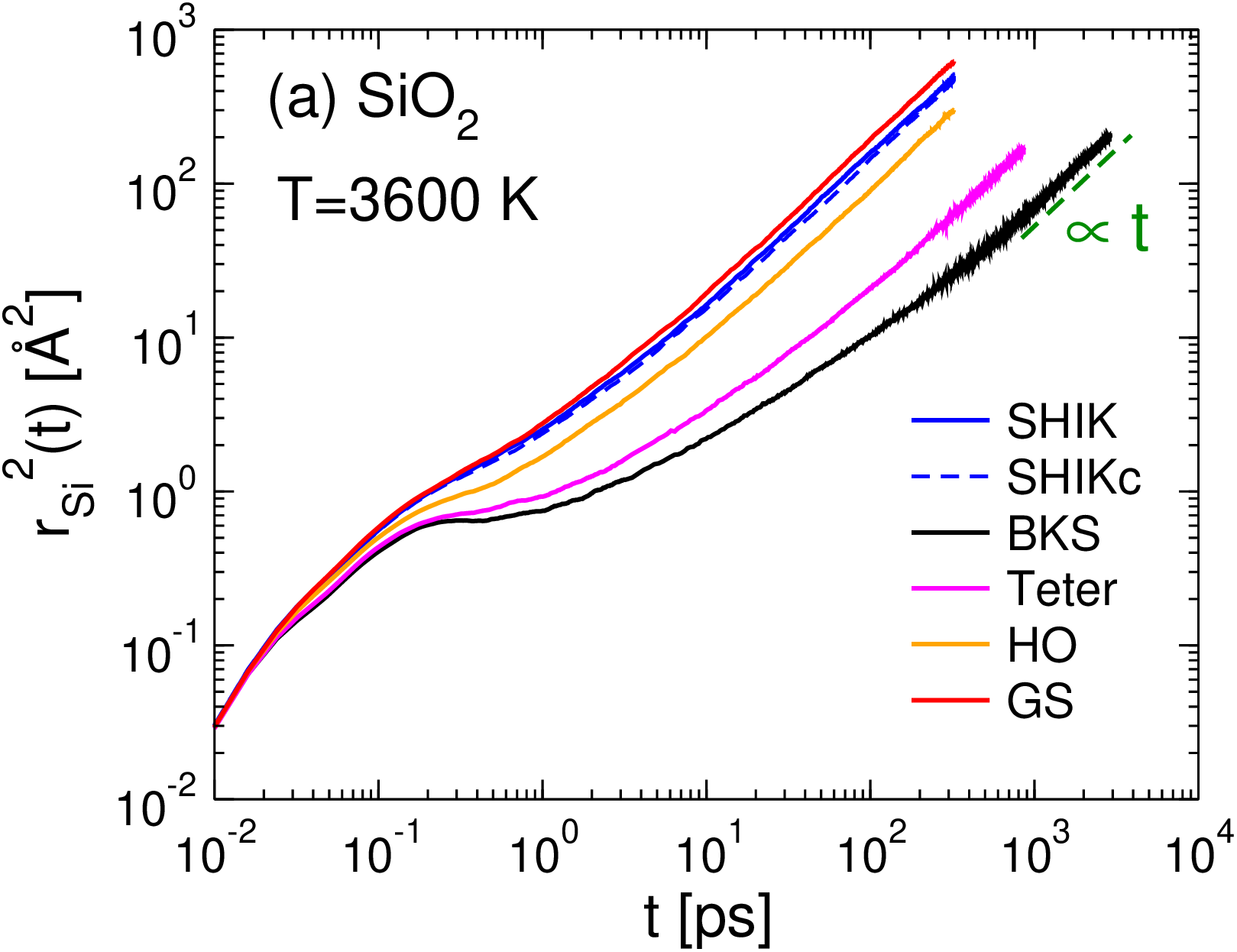}
\includegraphics[width=7.5cm]{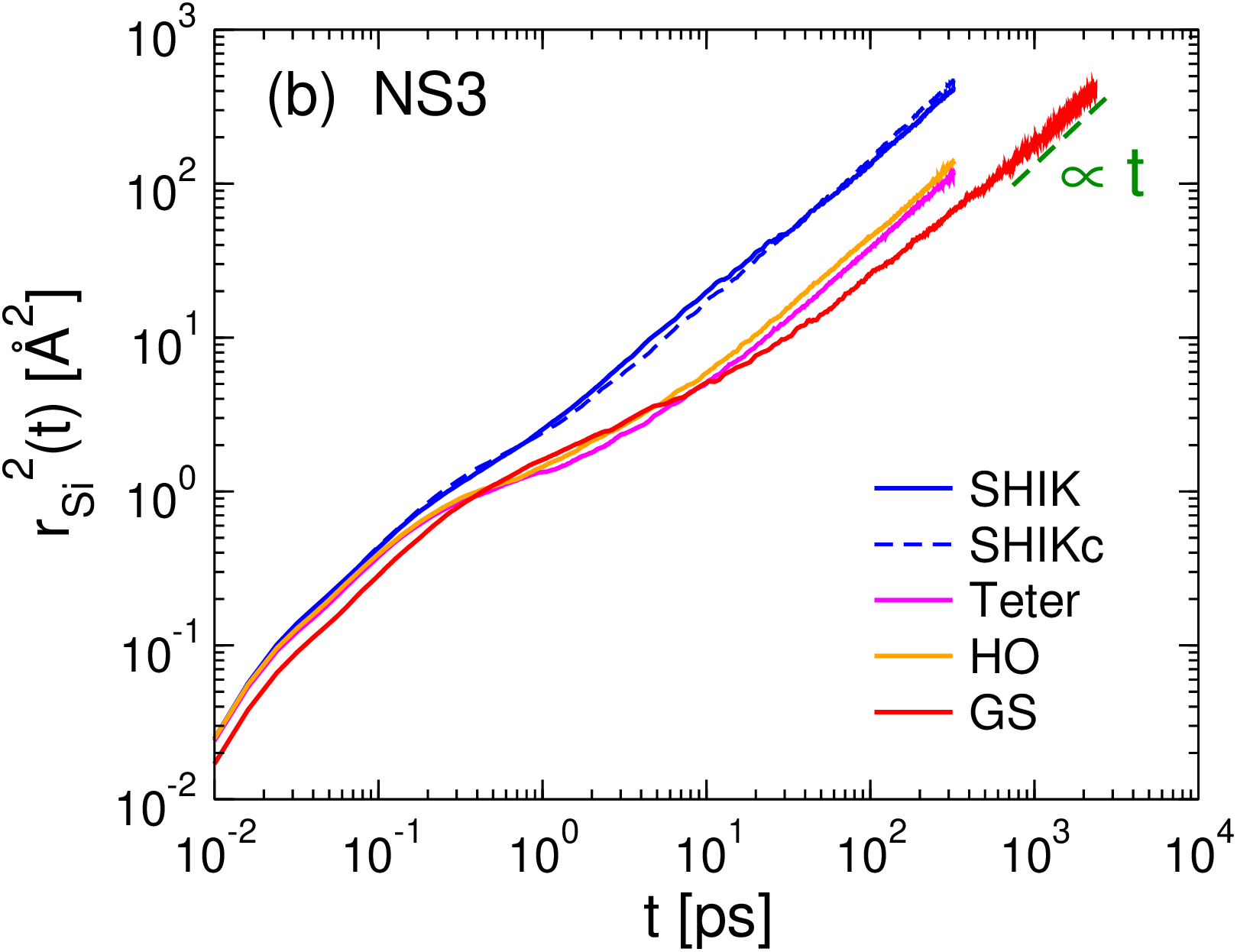}
\caption{
 Mean squared
displacement for silicon in the silica (a) and the NS3 (b) samples during
the $NPT$ equilibration at high temperatures and at zero pressure. The different curves correspond to the potentials used and the dashed line
shows the diffusive behavior.
For SiO$_2$ the temperature is 3600~K. For NS3 the temperature is 3000~K except for the GS curve which is for 2100~K.
}  
\label{fig:msd-si-liquid} 
\end{figure}

In Fig.~\ref{fig:msd-si-liquid} we show the time dependence of the MSD for
Si, i.e., the species that moves the slowest. The temperatures are those at
which we have equilibrated the samples in the $NPT$ ensemble and the curves
correspond to the different potentials. One sees that, at these $T$s,
the dynamics is already somewhat glassy in that at intermediate times the
MSD has a plateau~\cite{binder_glassy_2011}. These graphs also show that at long
times the MSD is a linear function of time, i.e., that the particles have
reached the diffusive regime, indicating that the system is equilibrated.

In agreement with previous results~\cite{hemmatti_comparison_2000}, we find that
the MSD at long times, and hence the diffusion constant, shows a very
strong dependence on the potential, more than a factor of ten, although
the structure of the glass does not vary that much (see below). Since
it can be expected that the activation energy for the diffusion constant
also depends on the potential considered, the diffusion constant at
lower temperatures will differ even more, as previously reported in Ref.~\cite{hemmatti_comparison_2000}.

\begin{figure}[htbp]
\centering
\includegraphics[width=7.5cm]{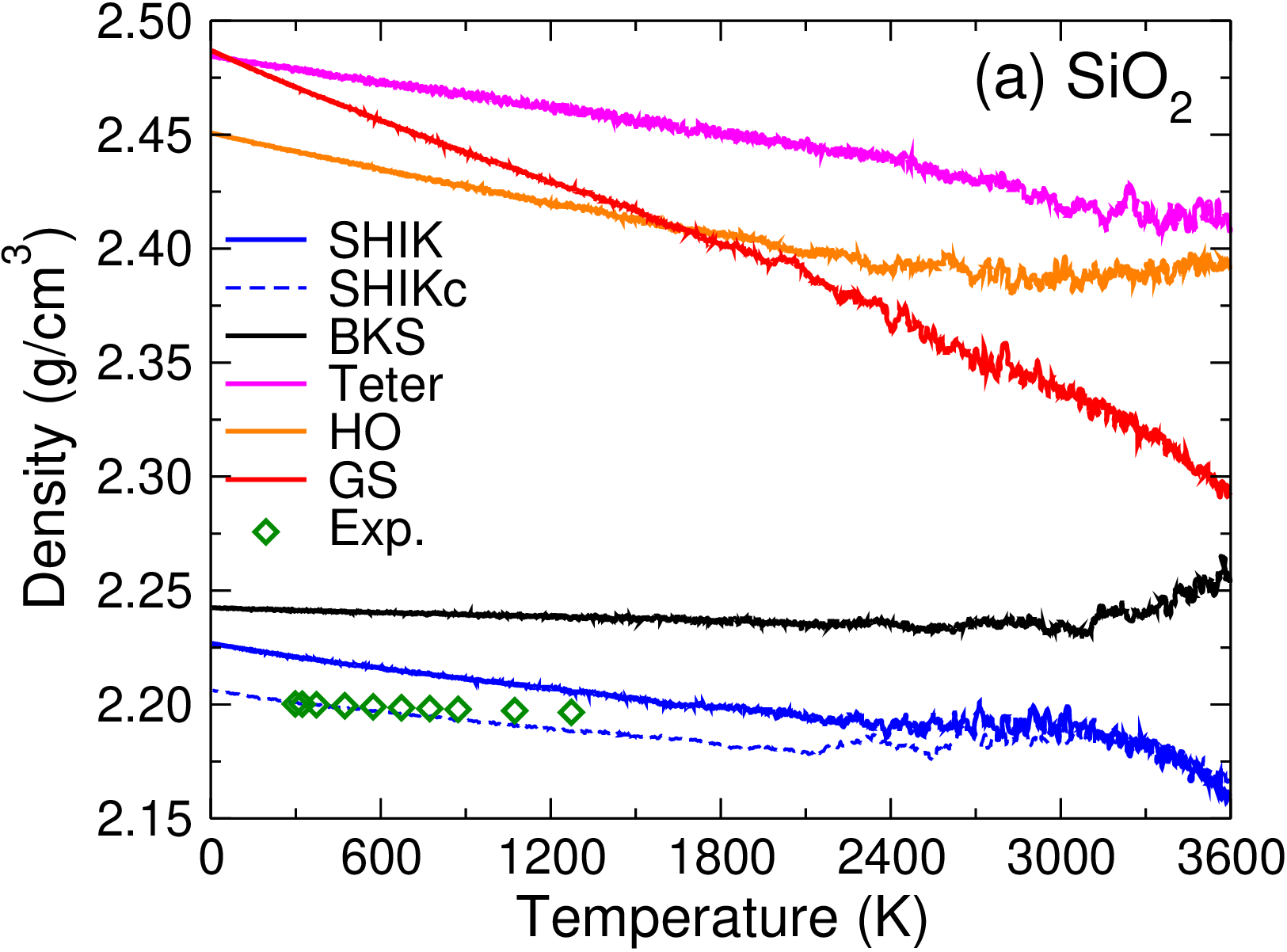}
\includegraphics[width=7.5cm]{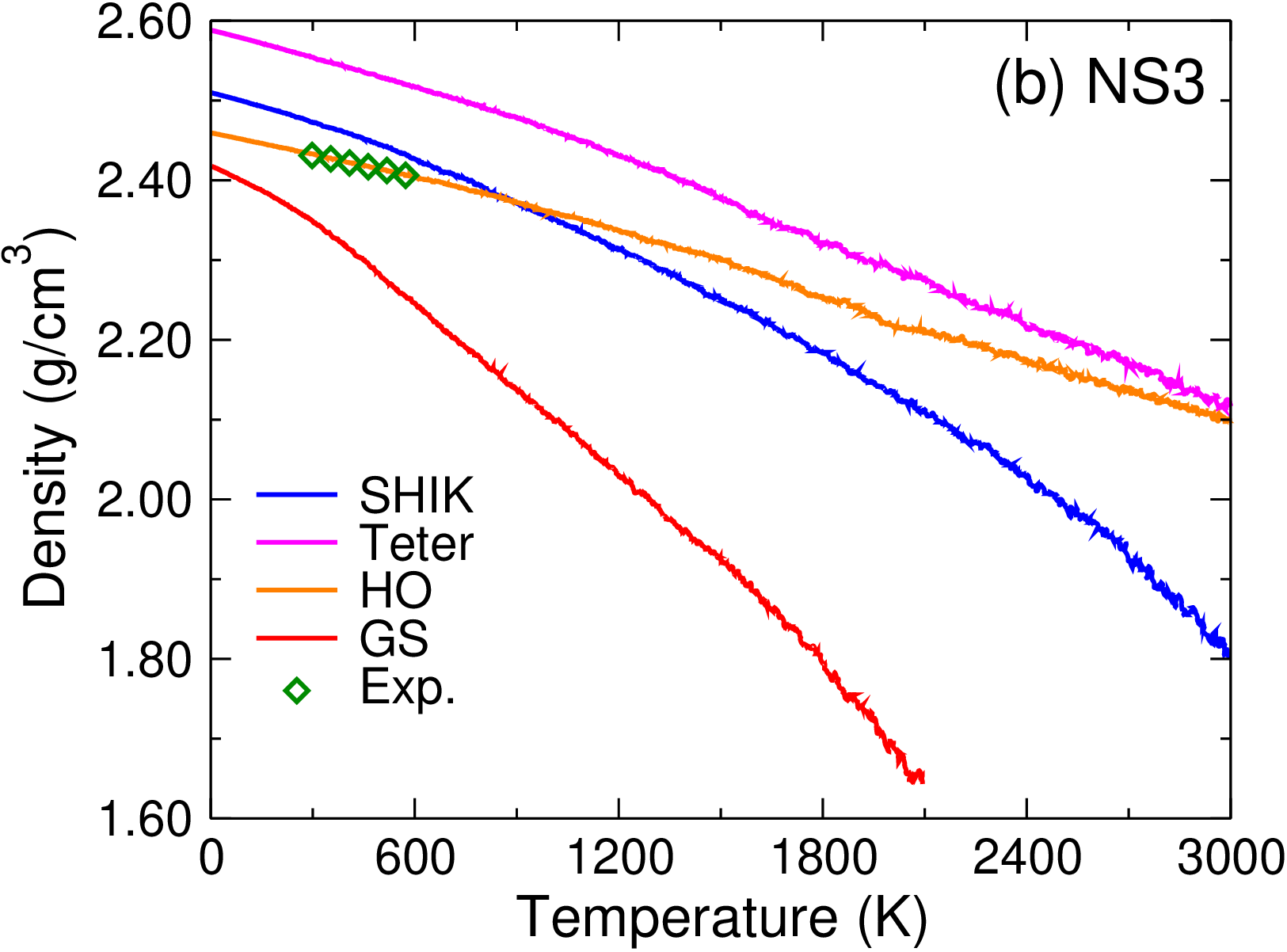}
\caption{
\label{fig:rho-quench} Temperature dependence of the density  during the quench at zero pressure for silica and NS3, panels (a) and (b), respectively.
}
\end{figure}

In Fig.~\ref{fig:rho-quench} we show the temperature dependence of the
mass density as predicted by the various potentials. These curves were
obtained by cooling the samples at zero pressure from high temperatures
to 0~K, i.e.~at high $T$ they correspond to the equilibrium density of
the system whereas at low $T$ to the density of the glass phase.
The data demonstrates that the $T-$dependence of the density depends
significantly on the potential: Not only the absolute values differ but
also the slopes, i.e.~the thermal expansion coefficients. For NS3 these
slopes are larger than the ones for SiO$_2$ a result that matches the
experimental findings~\cite{bansal_handbook_1986}.  Also included in the
graphs are the experimental values of the density at room temperature
(left most green diamond) and higher.  For  SiO$_2$, one sees that
 three potentials, namely Teter, HO, and GS, predict a density that
is significantly too high, whereas the BKS and SHIK potentials are
able to predict this quantity quite well. For NS3, the HO and SHIK
potentials  make an accurate prediction of the experimental density
(green diamonds) whereas the ones by Teter and GS predict values that
differ somewhat more from the experimental one. These graphs show
thus that there are potentials which are able to give a quantitative
good prediction of the density at room temperature.  Note that the
densities at room temperature are influenced by the cooling rate of the
sample~\cite{vollmayr_cooling-rate_1996}. However, this dependence is
relatively mild since the density is directly related to the fictive
temperature at which the sample falls out of equilibrium and this
temperature depends only logarithmically on the cooling rate. Thus if the
discrepancy between the predicted density and the experimental value is
too large, this cannot be rationalized by a too high cooling rate but
must instead be considered as a flaw of the used potential.

We have also included  in Fig.~\ref{fig:rho-quench} the
experimental values of the density at higher temperatures,
calculated using the  experimental values of the thermal expansion
coefficient~\cite{bansal_handbook_1986}. One recognizes that, for silica,
the slope of this data matches very well the one predicted by the SHIK
and BKS potentials while the three other potentials predict a larger
expansion coefficient. For NS3 the graph shows that the HO potential
predicts a slope which is in excellent agreement with the experimental
data, the SHIK and Teter potentials are in fair agreement, while the GS
potential is far off from the reality.

\begin{figure}[htbp]
\centering
\includegraphics[width=7.5cm]{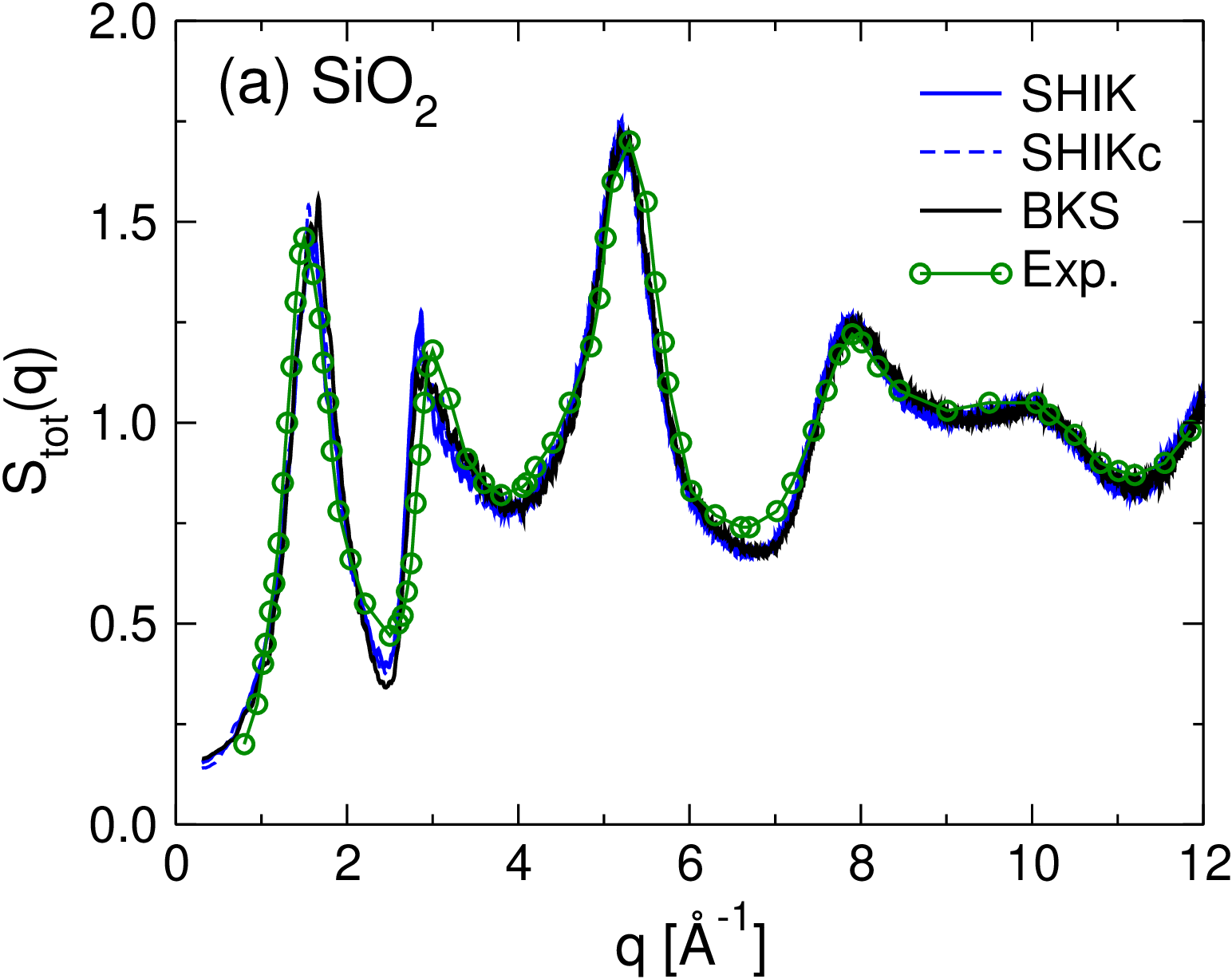}
\includegraphics[width=7.5cm]{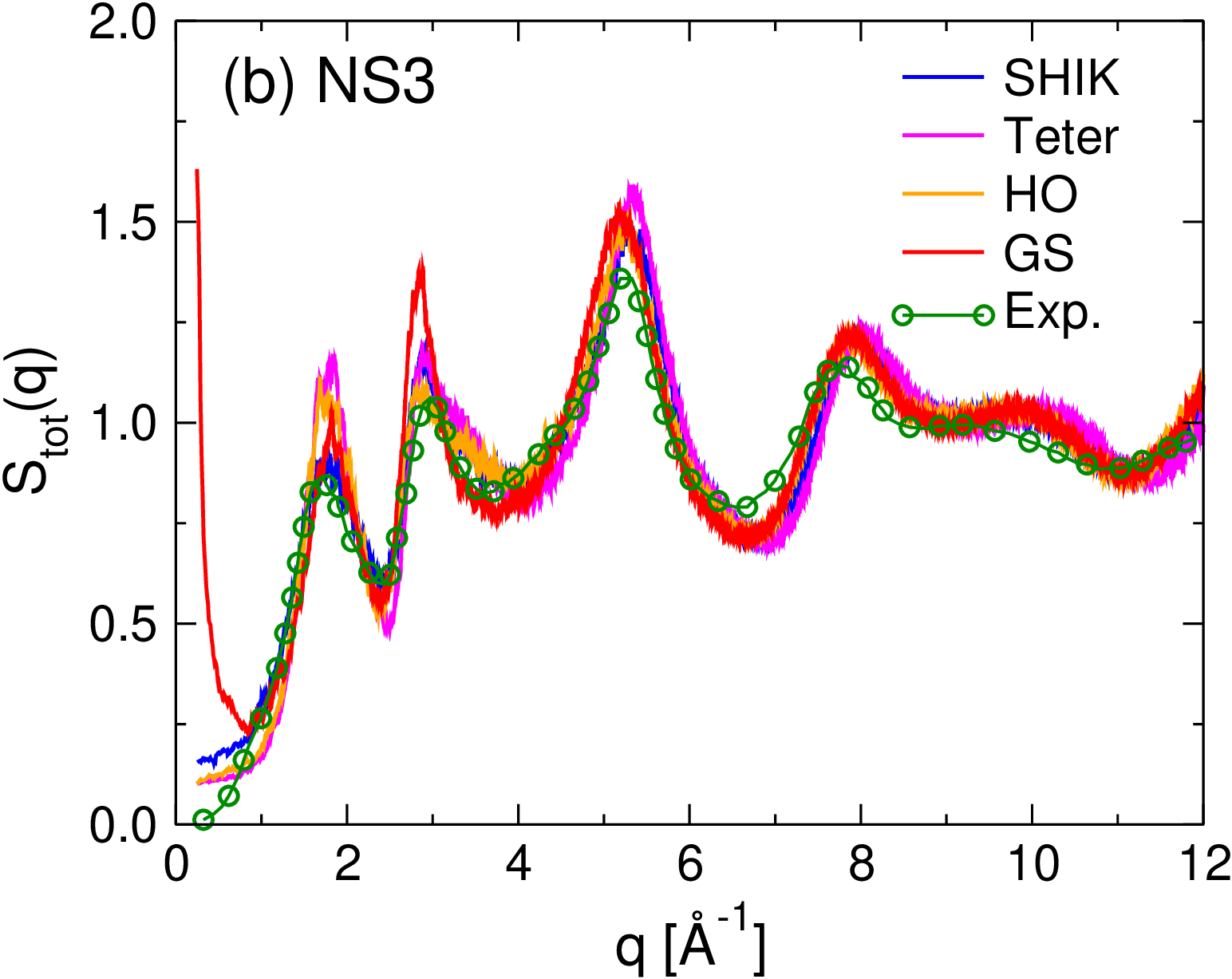}
\caption{
 Structure factor
as seen in neutron scattering experiments for silica (a)
and NS3 (b) glasses at 300~K.  Experimental data for silica are taken
from Ref.~\cite{susman_temperature_1991} and for NS3 from Ref.~\cite{pohlmann_structure_2005}.
}
\label{fig_structurefactor}
\end{figure}

We now probe how the properties of the glass depend on the potential. To
start we show in Fig.~\ref{fig_structurefactor} the static structure factor
as seen in a neutron scattering experiment which is the weighted sum of
the partial structure factors~\cite{binder_glassy_2011}:

\begin{equation}
S_{\rm tot}(q) = \frac{N}{\sum_{\alpha=\mathrm{Si, O, Na}} N_\alpha b_\alpha^2}
\sum_{\alpha, \beta=\mathrm{Si, O, Na}}  b_\alpha b_\beta S_{\alpha \beta} (q)
\quad .
\label{eq:structure_factor}
\end{equation}

\noindent
Here $q$ is the wave-vector, $b_\alpha$ is the neutron scattering length
for species $\alpha$ and $S_{\alpha\beta}(q)$ is the partial structure
factor: $\displaystyle S_{\alpha\beta}(q) =  \frac{f_{\alpha\beta}}{N} 
\sum_{j=1}^{N_\alpha} \sum_{k=1}^{N_\beta} 
\left \langle \exp(i{\bf q}.({\bf r}_j - {\bf r}_k)) \right \rangle$, 
$ \alpha,\, \beta = {\mathrm{Si, O, Na} }$, with  f$_{\alpha\beta}=1$ for ${\rm \alpha=\beta}$ and
f$_{\alpha\beta}=1/2$ otherwise, and $N$  the  total number of atoms. 
From the figure we recognize that for the case of silica all
considered potentials agree very well with the experimental data. This is
not that surprising since in most cases the parameters of these potentials
have been optimized to reproduce the structure of the glass. Qualitatively
the same conclusion can be drawn for the case of NS3, panel (b). There
is, however, one exception: The GS potential shows  a strong
increase of the signal at small $q$. This behavior indicates the presence of a phase
separation, in this case the formation of large domains of Na atoms, and
a visual inspection of the sample shows that this is indeed the case. (We
note that this defect of the potential is not readily seen in the radial
distribution functions and seems to have gone unnoticed so far.)

\begin{figure}[htbp]
\centering
\includegraphics[width=15cm]{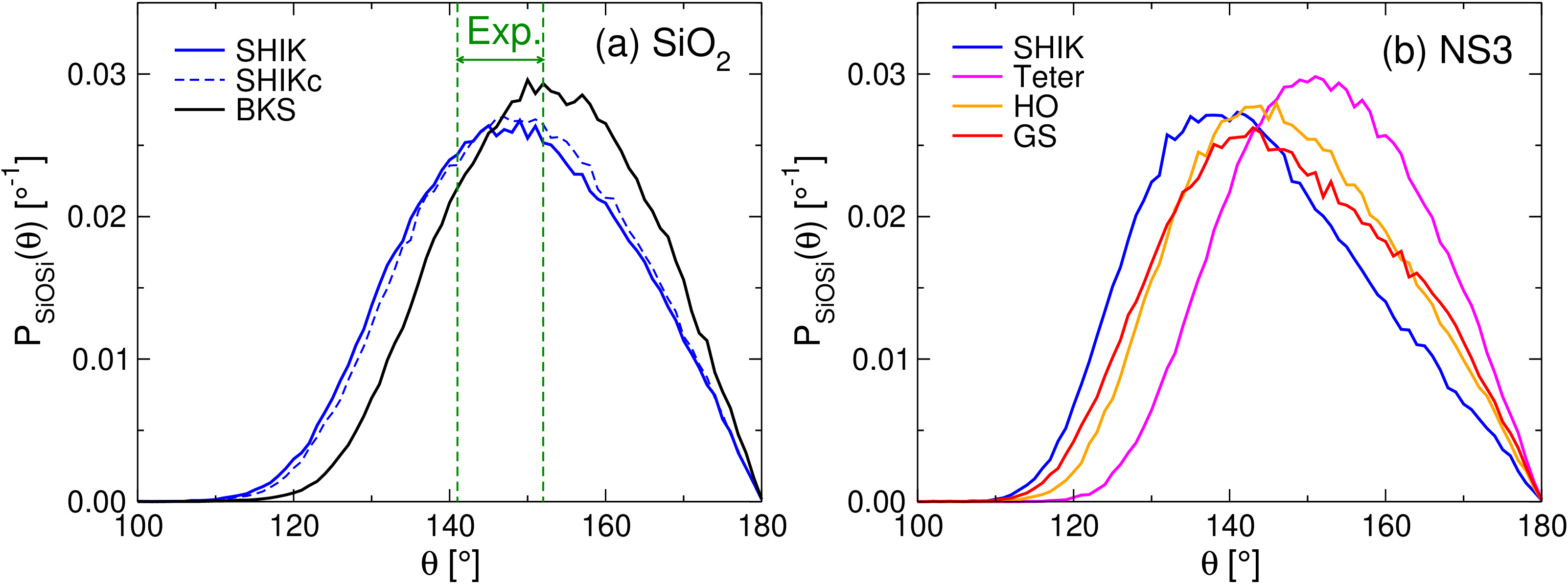}
\caption{
\label{fig:bad} 
SiOSi angle distribution for silica (a) and NS3 (b) glasses 
	at 300~K.
 The
vertical green lines indicate the range of most probable SiOSi angle
extracted from experimental studies (see Ref.~\cite{malfait_29si_2008}).
}
\end{figure}

A further useful quantity to characterize the structure of silicate
glasses is the distribution of the bond angles. Of particular
interest is the angle SiOSi since it gives information about the
relative orientation of two neighboring tetrahedra. For the case
of SiO$_2$ this distribution is presented in Fig.~\ref{fig:bad}a
and we recognize that there is not much dependence on the potential
considered and that all of them predict a position of the maximum of
the distribution that is compatible with the experimental estimate for
that angle~\cite{malfait_29si_2008}. Also for NS3 the distribution
of this angle is basically independent of the potential used,
Fig.~\ref{fig:bad}b. The only exception is the data obtained from the
Teter potential which predicts a peak at significantly larger angles. The
calculated mean SiOSi angles are 143.7$^{\circ}$, 147.7$^{\circ}$,
147.1$^{\circ}$ and 147.7$^{\circ}$, and 152.3$^{\circ}$ for the
SHIK, GS, HO and Teter potential, respectively. When compared to the
experimental mean value of 141.7$^{\circ}$ extracted from $^{29}$Si
MAS NMR measurements~\cite{angeli_insight_2011},  the SHIK data is in
good agreement, while GS and HO present a reasonable one.  In addition,
one can compare these distributions with the ones from {\it ab initio}
simulations~\cite{kilymis_vibrational_2019,tilocca_structural_2006}
of very  similar compositions, which predicted peak positions close to
140$^{\circ}$, i.e.~values that are compatible with the one predicted
by the SHIK, HO, and GS potentials, but not with the prediction by the
Teter potential.

\begin{figure}[tbp]
\centering
\includegraphics[width=15cm]{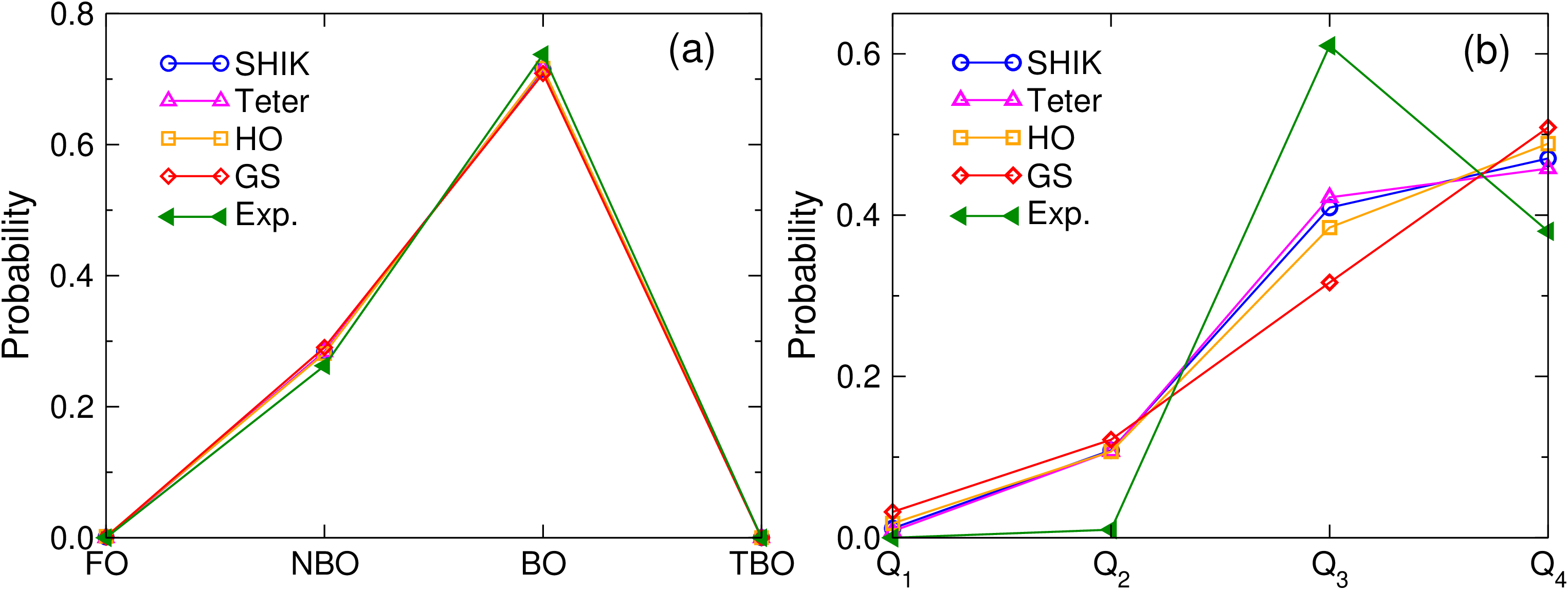}
\caption{Distributions of oxygen speciations, panel (a),  and
$Q_n$ species, panel (b), for the NS3 glass 
{at 300~K}. The experimental data
in panels (a) and (b) is from Ref.~\cite{nesbitt_bridging_2011}, and
Ref.~\cite{maekawa_structural_1991}, respectively.}
\label{fig:otype-Qn} 
\end{figure}

Finally we discuss two local structural quantities that probe the local
environment of an atom, namely the oxygen speciation and the distribution
of the $Q_n$ tetrahedral species in NS3.
 For the former we have used the first minimum
in the radial distribution function of the Si-O pair to determine the
connectivity of a given oxygen atom and thus to determine whether the atom
is free (FO), non-bridging (NB), bridging (BO), or three-fold coordinated (TBO).
In Fig.~\ref{fig:otype-Qn}a we show the corresponding  probability for
the different potentials and one recognizes that the considered force
fields all give the same distribution and it agrees very well with the
experimental measurements~\cite{nesbitt_bridging_2011}.
Thus one can conclude that this quantity is very robust or in
other terms the oxygen speciation is not very useful indicator for
testing the quality of a potential.  Things are different for the
$Q_n$ species, i.e.~the probability that a silicon atom is connected
to exactly $n$ bridging oxygens, $n = 0,1,2,3,4$. These probabilities are shown in
Fig.~\ref{fig:otype-Qn}b and one finds that, e.g., the frequency of the
$Q_3$ structure depends significantly on the potential. In particular
one observes that the GS potential gives a probability that is rather low
with respect to the other potentials and also to the experimental data.
	Thus the observation that the fraction of $Q_3$ units 
depends significantly on the used potential indicates that this quantity can be used as an indicator for evaluating the quality of a potential.
In addition we mention that the network depolymerization  depends also on the cooling rate: For sodium silicate glasses, it has been shown that the percentage of $Q_3$  increases with decreasing quench rate while the $Q_4$ one is decreasing, and hence improve the agreement with the experimental data~\cite{li_cooling_2017}.

\subsection{
\label{sec:tensile}
Mechanical properties of the glass}

To determine the mechanical properties of the annealed glass samples
we strained them in one direction using a strain rate of 0.5/ns. As we
will show below this value is sufficiently small to obtain results that
do not depend in a significant manner on the rate. 
We considered two
ensembles for this transformation: A constant pressure ensemble ($NPT$)
in which the pressures in the directions orthogonal to the strain were
set to zero and a constant volume ensemble ($NVT$) in which the two
orthogonal directions were not allowed to change, i.e.~the cross section
was constant. From this transformation we can thus directly obtain the
stress-strain curve in the pulling direction discussed in the following.

\subsubsection{System size and cooling rate dependence}
\label{sec:systemsize}

\begin{figure}[tbhp]
\centering
\includegraphics[width=7.5cm]{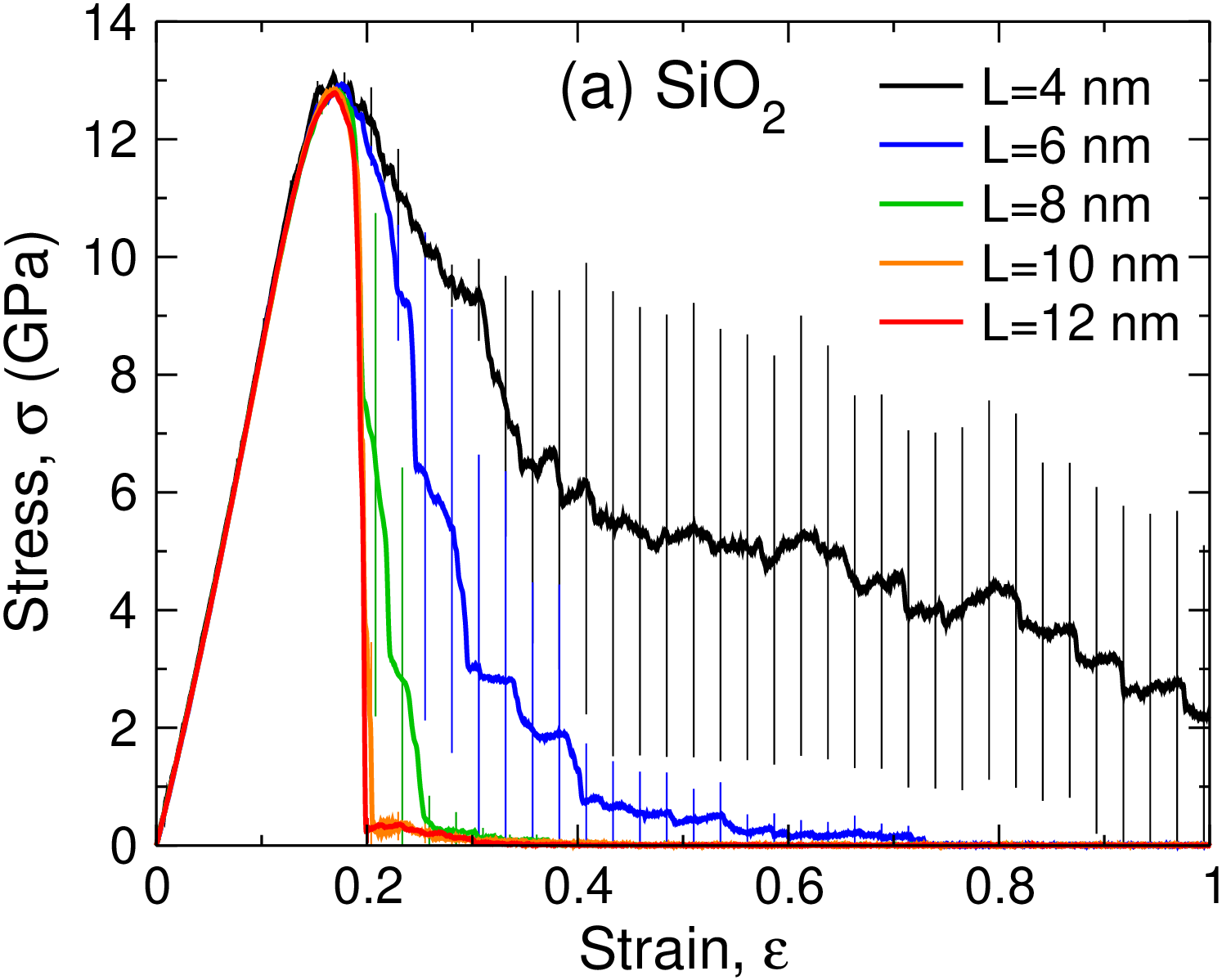}
\includegraphics[width=7.5cm]{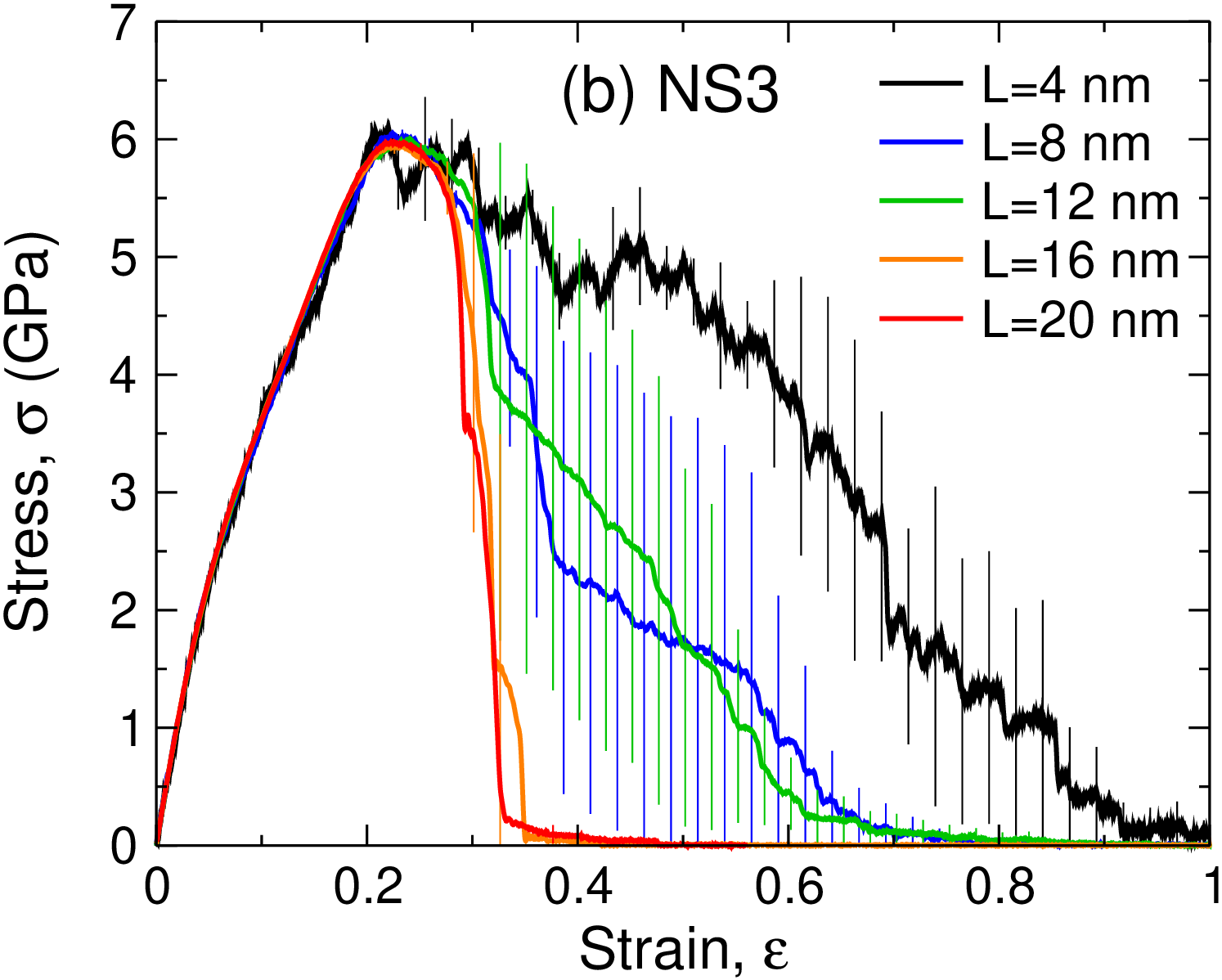}
\caption{Dependence of stress-strain curve on the system size for the
SHIK potential 
{at 300~K}. (a) Silica and (b) NS3.  The system size ranges from
around 5000 to around 600,000 atoms, corresponding to the increase of the
size of the simulation box size  from 4~nm to 20~nm at room temperature
density (at zero strain). The error bars were estimated from 
the standard deviation for the three samples.
}  
\label{fig:ss-systemsize}
\end{figure}

Before starting the discussion of the physical properties of the systems,
we probe how these properties depend on the system size. Usually the
properties of glasses can be obtained from simulations with relatively
moderate system size, say $10^3-10^4$ atoms. However, elasticity
and fracture are associated to  non-local processes and hence finite size
effects can be important. That this is indeed the case is demonstrated
in Fig.~\ref{fig:ss-systemsize} which shows the stress-strain curve
for different (cubic) box sizes $L$. (These results are for the SHIK
potential, but for the other force fields a similar behavior has been
found~\cite{zhang_fracture_2020}.) Error bars have been estimated by
considering three different samples.  For the case of silica, panel
(a), one finds that the elastic regime is basically independent of
$L$. However, once the failure point (i.e.~the strain at which the
stress has a maximum)  has been attained there are very strong finite
size effects in that the small systems break in a much gentler manner
than the large ones. Only if $L$ has reached 10~nm, which corresponds
to around 75,000 atoms, the stress-strain curve becomes basically
independent of the system size (see  Refs.~\cite{yuan_molecular_2012}
and \cite{pedone_dynamics_2015} for related studies).

For the case of NS3 the system size dependence is more pronounced in
that one has to use systems of about $L=16$~nm, corresponding to about
300,000 atoms, before the stress-strain curve converges, see panel
(b) in Fig.~\ref{fig:ss-systemsize}.  These  stronger finite size
effects are likely related to the presence of the Na atoms which make
a more heterogeneous structure in the NS3 glass compared to that of
SiO$_2$~\cite{greaves85,horbach02}.  However, for strains smaller than
the failure point the curves for the different system sizes superimpose
quite nicely and therefore also for this composition a system size with
around 36,000 atoms, i.e.~box size 8~nm, is sufficient to study the
elastic regime.

In order to make a fair comparison of the behavior predicted by the
different potentials we have used the same system size
$L=8~$nm, which corresponds to $N=36480$ and $N=38400$ atoms for SiO$_2$
and NS3, respectively. Although for this system size one still can
observe finite size effects, they are minor and hence do not preclude to
understand which potentials give rise to a realistic fracture behavior
and which ones do not.


\begin{figure}[tbhp]
\centering
\includegraphics[width=0.9\textwidth]{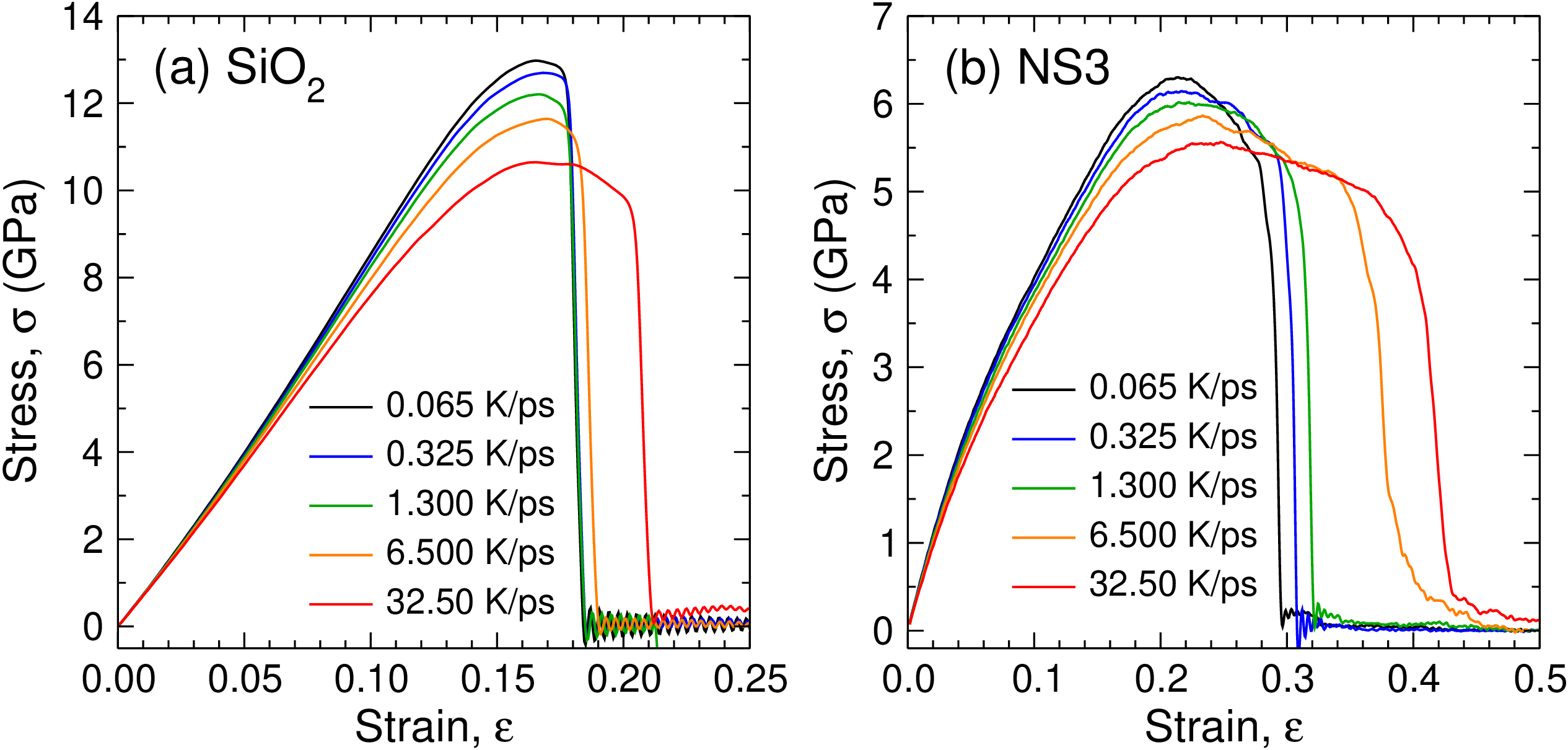}
\caption{ 
	Stress-strain curves as predicted by the SHIK potential for glasses at 300~K, produced with different cooling rates. Panels (a) and (b) are for silica and NS3, respectively.
}
\label{fig:cooling_rate_dependence}
\end{figure}

	Another important parameter for the properties of a glass is the cooling rate with which the sample has been produced. In the past it has been found that structural quantities show a clear dependence on this cooling rate, often described by a logarithmic law~\cite{vollmayr_cooling-rate_1996,lane_cooling_2015,li_cooling_2017} and hence one can expect that also the mechanical properties depend on this rate. In Fig.~\ref{fig:cooling_rate_dependence} we show that this is indeed the case in that with decreasing rate the failure strain decreases while the failure stress increases. At the same time a decreasing cooling rate leads to an increase of the slope in the elastic regime, i.e.~the system becomes stiffer. Although these cooling rate effects are noticeable, we see that once the rate is below 0.3~K/s the effect is minor and hence the results discussed below, obtained with a rate of 0.25~K/s, are only mildly affected by the cooling rate used to produce the glass sample. (Note that the samples used in Fig.~\ref{fig:cooling_rate_dependence} were produced using the SHIK potential and by quenching firstly the liquids in the NVT ensemble to a temperature a bit below the glass transition, and then to 300~K in the NPT ensemble~\cite{zhang_fracture_2020}. These samples contained 600,000 atoms.)

\subsubsection{Fracture behavior and elastic moduli}
\label{sec:fracture_moduli} 

\begin{figure}[tbhp]
\centering
\includegraphics[width=15cm]{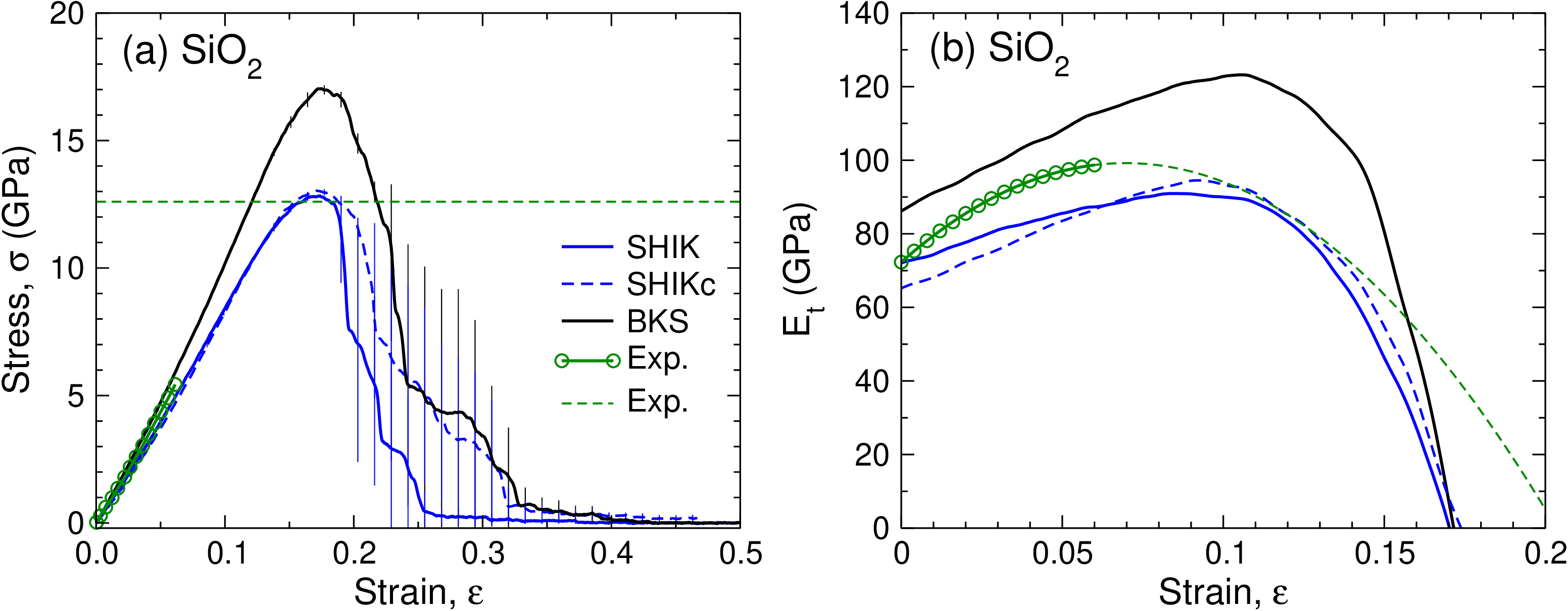}
\includegraphics[width=15cm]{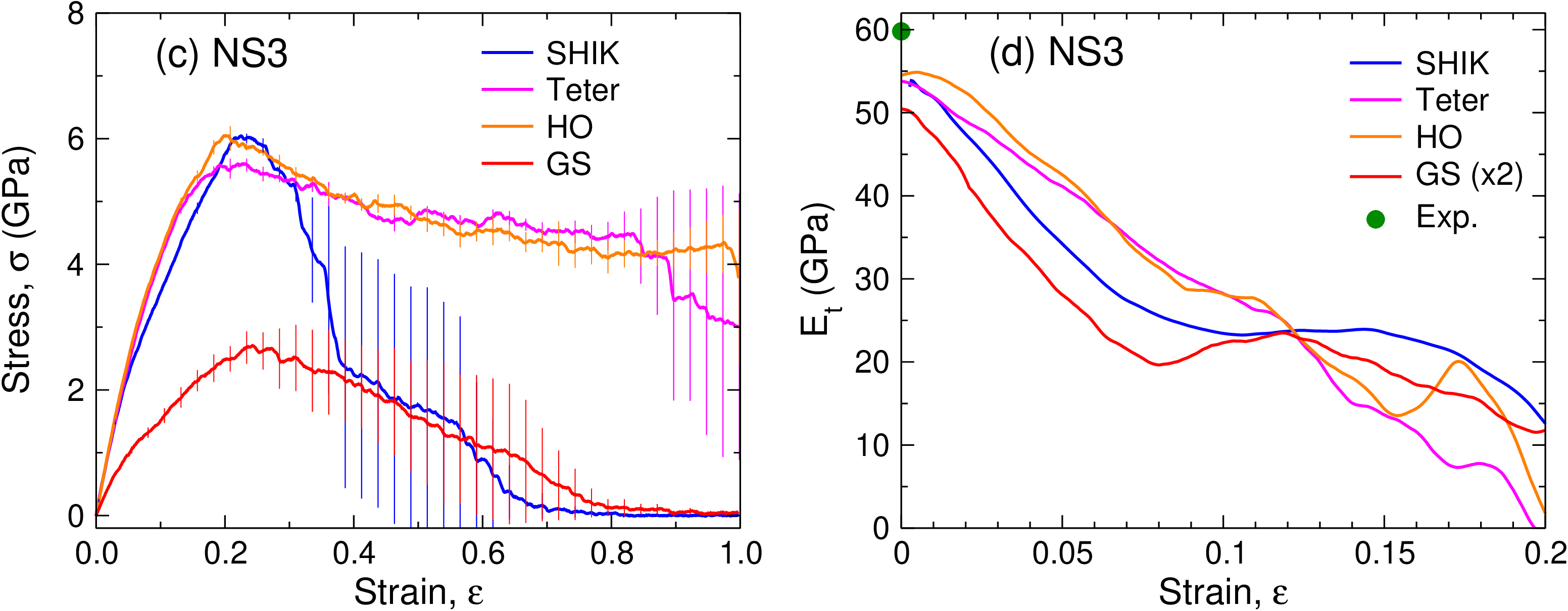}

\caption{Mechanical responses of the glasses under
uniaxial tension  
{at 300~K}. Panels (a) and (b) show the stress-strain curve and the
tangent modulus for silica, respectively. The green symbols are the experimental
data from Ref.~\cite{krause_deviations_1979}.
The green dashed line in panel (b) is the polynomial
fit from Ref.~\cite{gupta_intrinsic_2005} to experimental data and its
extrapolation to large strains. Panels (c) and (d): Same quantities as in panels (a) and (b) for the case of NS3.
}
\label{fig:ss-frac}
\end{figure}

We now discuss how the elastic moduli and the fracture behavior depend
on the potential, the ensemble, and the way the Coulomb interaction is
handled. In Fig.~\ref{fig:ss-frac}a we compare the stress-strain curves
for BKS with the one for SHIK (obtained in the $NPT$ ensemble).
One sees that the BKS glass is stiffer and stronger than the SHIK glass,
and also breaks in a more ductile manner.

Also included in the graph is the data for the SHIKc potential with the
full Coulomb interaction (blue dashed line). One sees that the results
are very close to the SHIK curve (blue full line), i.e.~when the Coulomb term is
replaced by the expression given in  Eq.~(\ref{eq:wolf}), and   hence
one can conclude that Wolf truncation is a good approximation, a result
which from the computational point of view is most advantageous. (Already
for 38400 atoms one gains a factor of around 3 in CPU time.)

From the stress-strain curve one can obtain directly the tangent modulus
which is defined as

\begin{equation}
E_t = \frac{d\sigma}{d\epsilon} \, ,
\label{eq:tang_mod}
\end{equation}

\noindent
where $\sigma$ is the stress in the pulling direction and $\epsilon$ the
strain. This quantity is shown in Fig.~\ref{fig:ss-frac}b for the various
potentials. In agreement with the curves from panel (a) we find that for
all strains the BKS glass has a tangent modulus that is significantly larger
than the one predicted by the SHIK potential. Interestingly, however,
both potentials predict the same failure strain, i.e.~$\epsilon_f
\approx 0.17$, in excellent agreement with the experimental value
of 0.18~\cite{griffioen_optical_1995}.  For the SHIK potential we have also included the
data as obtained by using the full Coulomb term and we see that the
corresponding curve (dashed blue line) agrees very well with the one in
which the Wolf term is used.

In addition, we have included in panels (a) and (b) of Fig.~\ref{fig:ss-frac}  the experimental data from Refs.~\cite{krause_deviations_1979} and \cite{gupta_intrinsic_2005}. One
sees that these data sets agree very well with the predictions of the SHIK
potential which is also able to reproduce accurately the Young's modulus, i.e.

\begin{equation}
E = \lim_{\epsilon\to 0} E_t^{NPT}(\epsilon) \, ,
\label{eq:young}
\end{equation}

\noindent
with a value equal to 72~GPa, in very good agreement with the experimental value given by 
73~GPa~\cite{bansal_handbook_1986}.  The prediction of the BKS potential is a bit
less satisfactory, 86~GPa, in agreement with the findings from Ref.~\cite{yuan_molecular_2012}.
For the failure strength the BKS and SHIK potentials predict 17.6~GPa and
12.8~GPa, respectively, and the latter value is in excellent agreement
with the experimental value of 12.6~GPa~\cite{griffioen_optical_1995}.

The stress-strain curve of NS3 as predicted by the different potentials is
shown in Fig.~\ref{fig:ss-frac}c). Surprisingly we find that under $NPT$
conditions the HO and Teter potentials  show no sign of fracture even
if the sample is stretched to 100\%. This shows that these potentials
have a serious flaw in that they strongly overestimate the ductile
behavior of NS3.  Also the stress-strain curve from the GS potential is
not realistic in that the stiffness in the elastic regime is strongly
underestimated, the failure stress is too small, and that the fracture
is way too ductile.  Hence we conclude that these three potentials
should not be used to study the fracture behavior of NS3.  A much more
reasonable stress-strain curves is found for the SHIK potential which
shows a relatively brittle fracture. This brittleness is, however, less
pronounced than the one found in SiO$_2$, see panel (a), in agreement
with the expectation that the addition of Na will make the glass more
ductile. The failure stress for the SHIK potential is around 6~GPa
(see Table~\ref{tab:moduli_strength} for exact values), which compares
reasonably well with the experimental values that are between 7.5 GPa and
11.7~GPa~\cite{lower_inert_2004,kurkjian_intrinsic_2001} (this
latter value is only an upper limit, see Ref.~\cite{lower_inert_2004}).

The Young's modulus of the NS3 glass, given by Eq.~(\ref{eq:young}),
is around 55~GPa for the SHIK, HO and Teter data (see
Table~\ref{tab:moduli_strength} for exact values), in good agreement
with the experimental value of 59.8~GPa~\cite{bansal_handbook_1986},
while  the GS potential  predicts a value of only 26~GPa, thus way
too small. A more notable difference between the various potentials
is found for the strain dependence of the tangent modulus, shown in
Fig.~\ref{fig:ss-frac}d: While for the HO and Teter potentials, $E_t$
decreases basically in a linear manner, the SHIK and GS potentials show at
intermediate strain a plateau before they decrease to zero. (Note that the
data for the GS potential has been multiplied by a factor of 2.0 in order
to bring it on the same scale as the other curves.) This plateau is also
directly visible in the stress-strain curves in that they show at around
$\epsilon=0.07$ a marked bend (see Fig.~\ref{fig:ss-frac}c). Elsewhere
we will show that this rapid change in the effective stiffness of the
sample is related to a change in the plastic behavior of the sample on
the atomic scale~\cite{zhang_fracture_2020}.

\begin{figure}[tbp]
\centering
\includegraphics[width=10cm]{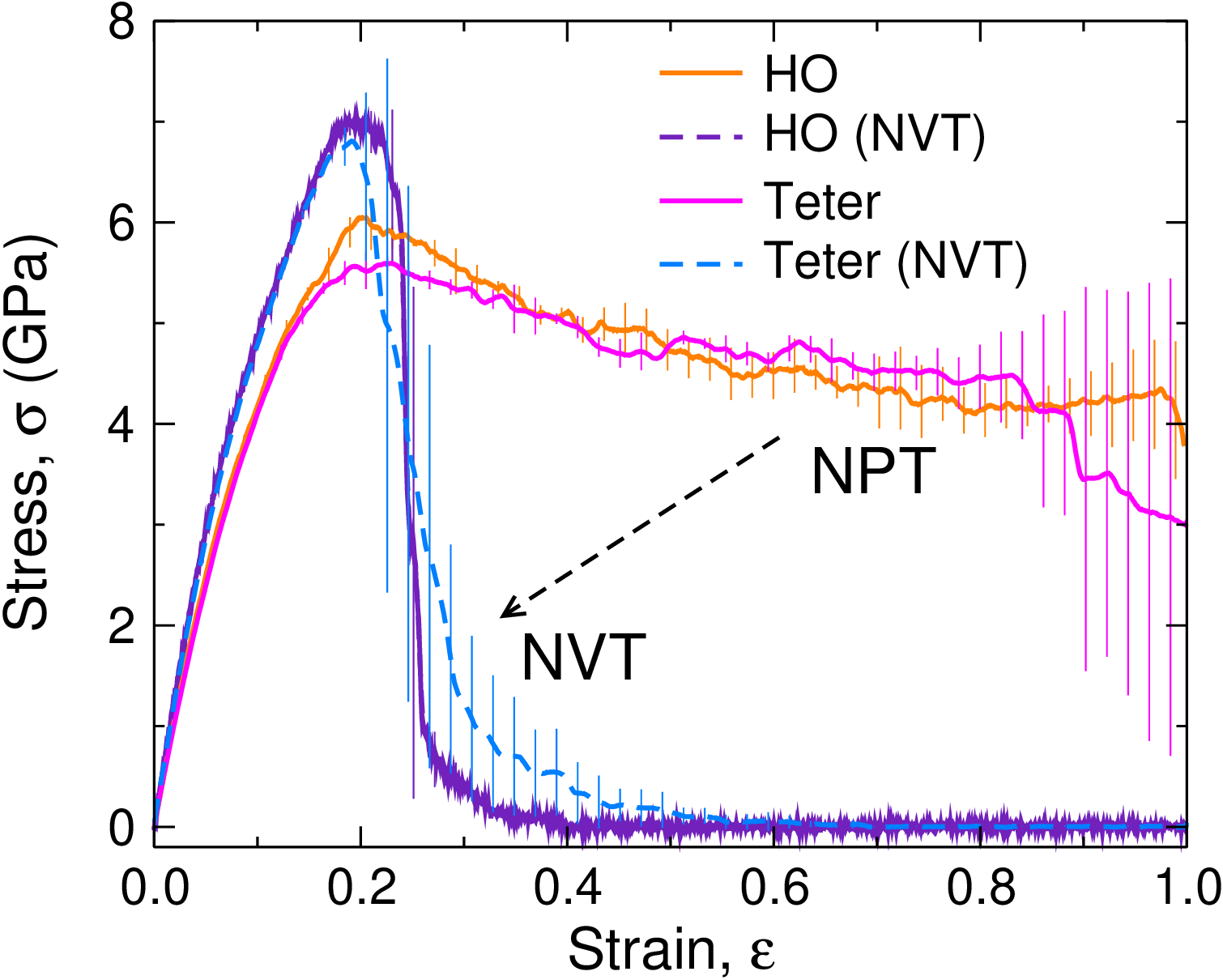}
\caption{Influence of simulation ensemble on fracture  
	{at 300~K}. The orange and
magenta curves
are the same as in Fig.~\ref{fig:ss-frac}c), i.e.~$NPT$
ensemble. The violet and blue dashed curves are the data obtained
in the $NVT$ ensemble. 
\label{fig:frac-ensemble}
}
\end{figure}

The fact that neither the Teter nor the HO potential predict  breaking if
strained to 100\% (see Fig.~\ref{fig:ss-frac}c), is astonishing since, from a structural point of view,  these potentials give reasonable predictions. Thus
one might wonder whether this unrealistic behavior is related to the
ensemble used during the tensile loading. In Fig.~\ref{fig:frac-ensemble} we
thus show the stress-strain curve for the two potentials but now in the
$NVT$ ensemble, i.e.~the sample size orthogonal to the pulling direction
is not allowed to change. We see that the so obtained stress-strain
curves are completely different from the ones in the $NPT$ ensemble in
that the formers indicate that the sample breaks in a rather brittle
manner. A similar ensemble effect has also been observed for the case of 
silica~\cite{pedone_molecular_2008}. Furthermore we note that in the elastic regime
the curve becomes somewhat steeper, i.e.~in the $NVT$ ensemble the system
is stiffer than in the $NPT$ ensemble, and this change is also seen for
the three other potentials, indicating that release of the constraint in
the orthogonal directions decreases the tangent modulus, a result that
certainly makes sense.

Finally we mention that,  for the BKS as well as SHIK potential for silica, the
fracture strain decreases significantly when one switches to the $NVT$
ensemble although the brittle nature of the fracture is independent of
the ensemble. (Note  that, as for NS3, the system is stiffer when working in the $NVT$ ensemble.)
From these results we thus conclude that the choice of
the ensemble is important, and thus  the $NVT$
ensemble should be avoided for this type of measurements.

\begin{table}[htbp]
	\centering
	\begin{tabular}{lllllllllll}
		\hline
		Quantity   &  & E  & C$_{11}$ & B & G & $\nu$  & $\epsilon_f$ & $\sigma_f$ & $\rho$  \\
		Unit   &  & GPa & GPa & GPa & GPa &   &  & GPa & g/cm$^3$ \\
		\hline
		\multirow{4}{*}{Silica} & SHIK     & 72.1   & 80.6  & 40.8    & 29.9    & 0.205 & 16.89  & 12.84  & 2.221 \\
		& SHIKc     & 65.0   & 76.0   & 40.9   & 26.3    & 0.235 & 17.31  &   13.01 & 2.200 \\
		& BKS   & 85.8   & 99.0      & 52.3    & 35.0    & 0.226 & 17.18  & 17.65 & 2.241 \\
		& Exp.     & 72.9\textsuperscript{a}    & 78.0\textsuperscript{a}      & 36.3\textsuperscript{a}    & 31.3\textsuperscript{a}    & 0.165\textsuperscript{a} & 18.00\textsuperscript{b}  & 12.6\textsuperscript{b}, 11-14\textsuperscript{c} & 2.201\textsuperscript{a} \\
		\hline
		\multirow{5}{*}{NS3}    
		& SHIK & 55.7   & 66.2      & 36.4    & 22.4    & 0.245 & 22.43  & 6.05 & 2.472 \\
		& HO & 54.8    & 68.1      & 39.3    & 21.6    & 0.268 & 19.99  & 6.05 & 2.433 \\
		& Teter    & 54.5    & 66.7   & 37.8  & 21.7    & 0.259 & 22.03  & 5.66 & 2.555  \\
		& GS       & 25.9    & 31.4  & 17.5    & 10.4  & 0.252 & 25.09  & 2.72 & 2.348 \\
		& Exp.      & 59.8\textsuperscript{a}, 56\textsuperscript{d}    & 69.5\textsuperscript{a}      & 37.2\textsuperscript{a}    & 24.3\textsuperscript{a}, 22\textsuperscript{d}    & 0.232\textsuperscript{a}, 0.25\textsuperscript{d} & 20.85\textsuperscript{e}  & 11.71\textsuperscript{e}, 7.5\textsuperscript{f} & 2.431\textsuperscript{a} \\
		\hline 
	\end{tabular}
	\caption{
		\label{tab:moduli_strength} 
		Elastic constants (Young's modulus $E$,  
		longitudinal modulus $C_{11}$, bulk modulus $B$), Poisson's ratio $\nu$, 
		failure strain $\epsilon_f$ and stress $\sigma_f$ (GPa), and density $\rho$ 
		at 300~K, compared with experimental data.
		\newline \textsuperscript{a}~Ref.~\cite{bansal_handbook_1986}
		\newline \textsuperscript{b}~Ref.~\cite{griffioen_optical_1995}
		\newline \textsuperscript{c}~Ref.~\cite{smith_doe_1990}
		\newline \textsuperscript{d}~Ref.~\cite{januchta2019elasticity}
		\newline \textsuperscript{e}~Ref.~\cite{lower_inert_2004} (strength at 77~K, and the failure stress is  overestimated,  see discussion section in Ref.~\cite{lower_inert_2004}).
		\newline \textsuperscript{f}~Ref.~\cite{kurkjian_intrinsic_2001} (strength at 77~K) 
	} 
\end{table}

In the linear elastic regime one can characterize the mechanical behavior
of the sample by the elastic constants given by the Young's modulus from 
Eq.~(\ref{eq:young}) and the longitudinal modulus given by

\begin{equation}
C_{11} = \lim_{\epsilon\to 0} E_t^{NVT}(\epsilon) \quad .
\label{eq:c11}
\end{equation}

\noindent From $E$ and $C_{11}$ one can then obtain the bulk modulus $B$,
the shear modulus $G$, and the Poisson's ratio $\nu$ using the following
relations~\cite{bansal_handbook_1986,zhao_situ_2012}:

\begin{equation}
E = \frac{9B(C_{11} - B)}{(3B + C_{11})}
\label{eq:bulkmod}
\end{equation}

\begin{equation}
G = \frac{3(C_{11} - B)}{4}
\label{eq:shearmod}
\end{equation}

\begin{equation}
\nu = \frac{E}{2G} - 1 \quad .
\label{eq:poissonratio}
\end{equation}

The resulting values for the different potentials as well
as the failure stress and failure strain are reported in
Tab.~\ref{tab:moduli_strength}. In this table we have also included the
experimental values and one recognizes that overall the SHIK potential
with the Wolf truncation is likely the most reliable potential. Note that
this potential has the merit to be applicable not only to NS3 but,
with the same set of parameters, also to SiO$_2$ and therefore also for compositions between the two systems~\cite{sundararaman_new_2019}.

\section{Summary and Discussion}

The goal of this work has been to investigate how the properties of
the glasses, in particular their mechanical behavior, depend on the
details of the interaction potential used to simulate the system, and
to also to study the influence of the ensemble in which the simulations
are done. In agreement with previous studies we have found that the
structural quantities show a relatively mild dependence on the interaction
potentials, with a notable exception for the GS potential that predicts
a phase separation into sodium rich/poor regions.  The temperature
dependence of the density as well as the distribution of the different
silicon species show instead a noticeable dependence on the potential
and hence these quantities are more sensitive indicators for evaluating
the quality of a potential.

A very strong dependence on the potential is found for the fracture
behavior of the sample: While the Teter and HO potentials predict a
plastic deformation at constant pressure even if the strain is 100\%, the
samples generated by the GS and SHIK potentials fail at a much smaller
strain. Therefore the two former potentials should not be used for this
type of studies. In addition, we have found that both structural and
mechanical properties are not realistic when using the GS potential. If
the sample is kept at constant geometry in the directions orthogonal to
the pulling directions all potentials predict a brittle fracture.

Our results demonstrate that certain structural quantities show
a dependence on the potential, but that the properties of the
fracture show a much stronger one. The latter dependence is, however,
definitely weaker than the ones found for dynamical quantities (see
Ref.~\cite{hemmatti_comparison_2000}, for example). This can be understood
from the fact that static properties depend mainly on the position of
the particles close to the local minimum of the potential energy, while
fracture and dynamic properties depend also on the position and height of
the saddle points connecting these minima. Since for many systems there
is only very limited experimental information about dynamic quantities,
while elastic and inelastic measurements are much more abundant in the
literature, the mechanical properties are thus a very attractive way to
probe whether or not a potential is reliable.

A further important result of our study is the dependence of the fracture
behavior on the system size. As in previous studies we have found that
the structural quantities do depend only in a mild manner on the size
of the system. In contrast to this, our study suggests that a reliable
description of the  fracture behavior can only  be obtained for systems
that have more than 75,000 particles, i.e.~a box size of the order
10~nm for the case of silica and $L=16$~nm (300,000~atoms) for NS3.
For smaller sizes, the elastic regime shows almost no dependence on the
system size, while this dependence becomes very strong after the failure
point and the system will present a fracture behavior that is much more
ductile than a sufficiently large system.

Finally we have also documented the fact that the fracture behavior
of the glass samples can show a strong dependence on the nature of the
boundary conditions. If the traction is done under constant cross section
condition, the samples will break in a brittle manner. However, if one
fixes the pressure, as it is the case as in an experimental set-up, some
potentials predict a much more ductile fracture or even no fracture at
all up to a strain of 100\%. This shows that for this type of studies
one should use the $NPT$ ensemble {\it and} a potential that gives a
realistic fracture behavior.

\section*{Acknowledgements}
Z.Z. acknowledges financial support by China Scholarship Council (NO. 201606050112).
This work was granted access to the HPC resources
of CINES under the allocation A0050907572
attributed by GENCI (Grand Equipement National de Calcul
Intensif).


\end{document}